\documentclass[10pt,prb,aps,floatfix,superscriptaddress,showpacs,footinbib,twocolumn,longbibliography]{revtex4-2}

\usepackage{graphicx,mwe}
\usepackage{xcolor}
\usepackage{bm}
\usepackage{bbm}
\usepackage[normalem]{ulem}
\usepackage{braket}
\usepackage[utf8]{inputenc}
\usepackage[T1]{fontenc}
\usepackage{amsmath,mathtools}
\usepackage{mathrsfs}
\usepackage{booktabs}       
\usepackage{longtable}
\usepackage{amssymb}
\usepackage{soul}
\usepackage{float}
\usepackage{pgfplots}
\pgfplotsset{compat=1.18}
\usepackage[colorlinks=true,urlcolor=blue,citecolor=blue,linkcolor=blue]{hyperref}
\def\umphys{
    Department of Physics, University of Michigan,
    Ann Arbor, Michigan 48109, USA
}
\def\uwphys{
Institute of Theoretical Physics, Faculty of Physics, University of Warsaw, Warsaw, Poland
}

\def\cas{
Beijing National Laboratory for Condensed Matter Physics and Institute of Physics, \\Chinese Academy of Sciences, Beijing 100190, China
}

\def\israel{
The Raymond and Beverley Sackler Center for Computational Molecular and Materials Science, Tel Aviv University, Tel Aviv 6997801, Israel
}

\def\telaviv{
School of Chemistry, Tel Aviv University, Tel Aviv 6997801, Israel
}

\def\telavivphys{
School of Physics, Tel Aviv University, Tel Aviv 6997801, Israel
}

\begin{document}
\title{Compact and Stable Representation of Real-Frequency Spectral Functions for Machine Learning
}
\author{Xinyang Dong}
\email{dongxy@iphy.ac.cn}
\affiliation{\cas}

\author{Ido Zemach}
\affiliation{\telavivphys}
\author{Lei Zhang}
\affiliation{\umphys}
\author{Guy Cohen}
\affiliation{\israel}
\affiliation{\telaviv}
\author{Emanuel Gull}
\affiliation{\umphys}
\affiliation{\uwphys}

\date{\today}

\begin{abstract}
We introduce a compact and stable moment representation for real-frequency Green's functions, hybridization functions, and self-energies for machine-learning applications, avoiding the inefficiency of dense frequency grids as well as the ill-posed analytic continuation of Matsubara approaches. The representation is constructed from Cayley-mapped trigonometric moments with the Jacobian included, which preserve spectral-weight normalization, tie the moment sequence to a positive matrix-valued spectral measure, and admit a systematic route to a pole representation via ESPRIT. This provides a fixed-dimensional learning target in which physical constraints such as normalization and positivity can be imposed directly.
Using a graph-attention neural network with FiLM conditioning, we benchmark the representation on single-orbital DMFT, antiferromagnetic DMFT, and a two-orbital impurity model.
The results demonstrate accuracy matching or exceeding that of direct frequency-domain learning, reliable reproduction of the density and staggered magnetization, stable self-energy reconstruction through Dyson equation inversion, and accurate recovery of matrix-valued spectra with orbital mixing.
\end{abstract}

\maketitle

\section{Introduction}

Spectral functions offer a common language for describing the excitations, optical response functions, and thermodynamic properties of solids and molecules. Because of their direct relation to Green’s functions, they occupy a central role in quantum many-body theories ranging from non-perturbative quantum embedding methods \cite{Georges_DMFT_1996,Kotliar_DFT_DMFT_2006} to perturbative many-body  approaches \cite{Onida02} for electronic structure and lattice models. In general, obtaining spectral functions requires the solution of an interacting quantum system.

Quantum embedding theories, such as dynamical mean-field theory (DMFT) \cite{Georges_DMFT_1996,Kotliar_DFT_DMFT_2006}, reformulate a correlated lattice or material problem in terms of an auxiliary quantum impurity model. 
In DMFT, this impurity model describes a local interacting subsystem coupled to a self-consistently determined noninteracting bath. 
Solving the DMFT equations then requires repeatedly computing the impurity Green's function or spectral function from the impurity parameters and bath hybridization. 
This non-perturbative impurity-solver step often becomes the computational bottleneck, especially at low temperatures, in multi-orbital systems, and in realistic materials calculations. 
These challenges have motivated growing interest in machine-learning approaches for Green's functions and spectral functions, including surrogate impurity solvers, data-driven DMFT schemes, and models for molecular and materials Green's functions \cite{Arsenault_MLAIM_2014,Sheridan_MLDMFT_2021,Lee_SCALINN_2025,Agapov_GreenNN_2024,Kakizawa_PINN_AIM_2024,Dong_equivGF_2024,Valenti_NNembedding_2026,Zhu_ML_2026,Sturm_MLAIM_2021,Ren_NRGspectralML_2021,Liu_impuritySpectrumDL_2024,Miles_KondoVAE_2021,Deng_RFDMFT_2025}. 
Existing machine-learning approaches broadly use either imaginary-axis Green's functions or real-frequency spectra as learning targets.

In the imaginary-axis formulation of quantum statistical mechanics, Green's functions are smooth, naturally represented on Matsubara-frequency grids, and numerically stable in many calculations. 
They can also be compressed using compact bases such as the Legendre \cite{Boehnke_Legendre_2011}, Chebyshev \cite{Gull2018}, intermediate-representation \cite{Shinaoka_IR_2017}, and discrete Lehmann \cite{Kaye_DLR_2022} bases, reducing storage and computational cost. 
However, obtaining spectral functions from imaginary-axis data requires analytic continuation, an ill-posed inverse problem that relies on additional regularization assumptions \cite{Jarrell_AC_1996,Fei_Nevanlinna_2021,Fei_Caratheodory_2021,Shao_SAC_2023,Zhang_minipole_2024_a}. 
This difficulty is particularly relevant for machine-learning predictions, where small errors in Matsubara Green's functions can be amplified in the resulting real-frequency spectra \cite{Zhu_ML_2026}.

Alternatively, one may learn spectral functions directly on the real-frequency axis, thereby avoiding analytic continuation altogether. 
Recent real-frequency machine-learning works have demonstrated the promise of this direction \cite{Sturm_MLAIM_2021,Ren_NRGspectralML_2021,Liu_impuritySpectrumDL_2024,Miles_KondoVAE_2021,Deng_RFDMFT_2025}. 
However, representing spectra on an equidistant real-frequency grid can be inefficient for machine-learning purposes, since dense grids contain discretization-dependent redundancy and neighboring frequency points are strongly correlated by the underlying analytic structure. 
Reducing this redundancy helps the model learn the spectral object itself rather than a large collection of grid values.

Compact and stable representations of real-frequency correlation functions are therefore important for scalable machine learning. 
In this manuscript, we introduce a moment-based representation for real-frequency Green's functions, spectral functions, and hybridization functions designed for machine-learning applications. 
Physical constraints, including normalization and positivity, can be imposed directly in moment space. 
At the same time, the finite moment sequence remains compact while retaining a systematic route back to the corresponding real-frequency spectrum.
We demonstrate the moment representation using quantum-impurity benchmarks of increasing complexity, including single-orbital DMFT on the Bethe lattice, symmetry-broken antiferromagnetic DMFT, and a two-orbital impurity model with matrix-valued spectra.
These benchmarks evaluate moment-space learning, real-frequency spectral reconstruction, and the extension of the representation from scalar to matrix-valued Green's functions.

\section{Methods}

\subsection{Spectral Representation} \label{sec:spec}
A representation suitable for machine learning of spectral functions should satisfy three requirements. First, it should respect the physical constraints of spectral functions by design, in particular positivity and normalization, thereby reducing the risk of unphysical predictions.
Second, it should be compact and accurate, limiting the number of degrees of freedom that need to be learned.
Third, it should be stable in the sense that small changes in the spectrum lead to controlled changes in the representation, and small errors in the representation do not produce large errors in the reconstructed spectrum.
This stability helps avoid the amplification of noise and prediction errors in the machine-learning workflow.

A parametrization in terms of a real-frequency grid does not satisfy (2)
and requires (1) to be imposed exactly. 
Standard techniques used in Matsubara calculations, such as the IR \cite{Shinaoka_IR_2017}, Chebyshev \cite{Gull2018}, Spline \cite{Kananenka2016}, Legendre \cite{Boehnke_Legendre_2011}, or Discrete Lehmann Representation \cite{Kaye_DLR_2022} bases, are very compact but inaccurate in practice when applied to real spectra: while they provide precise imaginary-axis properties, they typically cannot represent real-frequency functions accurately due to the ill-conditioned analytic continuation kernel.

In contrast, a recently developed minimal pole representation \cite{Zhang_minipole_2024_a} and its matrix-valued generalization \cite{Zhang_minipole_2024_b} provide a powerful compact representation on the real axis. Originally developed for analytic continuation of Matsubara data, the methods have since been applied to the representation of spectral functions as a sum of poles in the complex plane \cite{Zhang_minipole_2025}, as well as for time-extrapolation \cite{Erpenbeck26} and Matsubara diagrammatics \cite{Gazizova24}. 

A complex pole representation of the retarded Green's function is 
\begin{figure}[tb]
    \centering
    \includegraphics[scale=1]{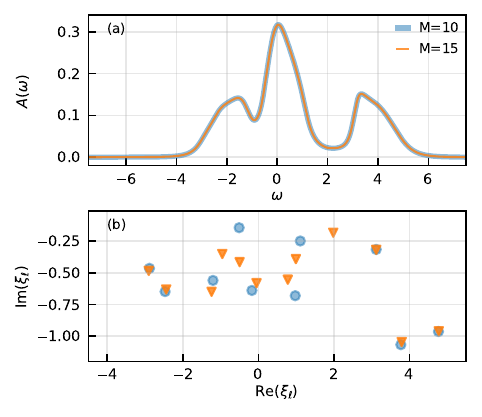}
    \caption{Ambiguity of the pole representation. (a) Spectral functions recovered with 10 and 15 poles are nearly indistinguishable, yet (b) the corresponding pole positions $\xi_\ell$ differ substantially, illustrating that the pole representation is not uniquely determined by the spectral function within the precision considered here.}
    \label{fig:poles}
\end{figure}
\begin{align}
    \mathbf{G}^\mathrm{R}(z)&\approx \sum_{\ell=1}^{M}\frac{\mathbf{A}_\ell}{z-\xi_\ell},
    \label{eq:pole_rep}
\end{align}
where the poles $\operatorname{Im}\xi_\ell<0$ lie in the lower half of the complex plane and the residues $\mathbf{A}_\ell$ are chosen such that the Nevanlinna-Herglotz properties of the retarded Green's function are satisfied \cite{Fei_Nevanlinna_2021,Fei_Caratheodory_2021}. While  Green's functions of lattice and impurity model systems typically require around 10 complex poles for their description to within numerical precision \cite{Zhang_minipole_2024_a}, substantially more may be needed to resolve structured Green's functions of transport systems \cite{Zhang_minipole_2025}.

In this representation, (1) and (2) are clearly fulfilled by construction. 
Fig.~\ref{fig:poles} examines (3): the stability of the pole locations as the number of poles is varied. 
It is evident that pole sets with different locations, orderings and residues can yield spectral functions that are identical on the real axis to within the target precision. 
Although strict uniqueness is not required for supervised learning, a useful learning target should be well conditioned and avoid artificial label ambiguity. 
Direct pole coordinates therefore do not satisfy this requirement and, while efficient and accurate for spectral reconstruction, are ill-suited as machine-learning targets.

\subsection{Moment Representation} \label{sec:mom}
Moment theory \cite{Akhiezer_moment_1965} provides an alternative representation. Moment representations are not as compact as pole representations but provide  stability and a way to enforce positivity and normalization by construction.
Moments of Green's functions have been explored in several contexts over the years, from commutator expressions characterizing high-frequency coefficients in Fourier transforms \cite{Comanac07},  the construction of effective Hamiltonians by moment matching \cite{Rusakov14},  reformulations of embedding theories in terms of moment problems \cite{Backhouse22}, mathematical analyses of perturbation theories \cite{Farid21} and high energy physics \cite{Abbott26}.

A positive spectral measure, such as a spectral function or (up to a scaling factor) the spectral density associated with a hybridization function or the dynamical part of a self-energy, can be characterized by a sequence of moments
$\hat{\mathbf{m}}_k=\int \mathbf{A}(\omega)\,\omega^k d\omega$.
A numerical complication arises from the rapid growth of higher-order moments in systems with spectral weight at large frequencies, making the naive parametrization in terms of moments on the real axis tedious in practice.

It is therefore more convenient to work with the so-called trigonometric moments \cite{Geronimus_trigono_1946} of a transformed problem instead. Using the Cayley transform and its inverse,
\begin{align}
    u = f(z)&=\frac{z+i\omega_p}{z-i\omega_p},
    \quad
    z = f^{-1}(u)=i\omega_p\frac{u+1}{u-1},
    \label{eq:cayley}
\end{align}
we map the real axis onto the unit circle  in analogy to Ref.~\onlinecite{Zhang_minipole_2025}. The free parameter $\omega_p>0$ sets the characteristic frequency scale of the mapping. We define the trigonometric matrix moments as
\begin{align}
    \mathbf{m}_k&=
    \int_{-\infty}^{\infty}d\omega\,
    \mathbf{A}(\omega)\,[f(\omega)]^k ,
    \label{eq:mom_w}
\end{align}
in which we include the Jacobian of the transformation. In this convention, the zeroth moment is the physical spectral weight
\begin{align}
    \mathbf{m}_0&=\int_{-\infty}^{\infty}d\omega\,\mathbf{A}(\omega).
    \label{eq:m0_sum}
\end{align}
Due to normalization, $\mathbf{m}_0 =\mathbbm{1}$ for Green's functions in orthogonal bases. Since $\mathbf{A}(\omega)$ is a positive matrix-valued spectral measure, its trigonometric moments satisfy a positivity condition: the block-Toeplitz matrix 
\begin{align}
    \mathcal{T}_N&=
    \begin{pmatrix}
        \mathbf{m}_0 & \mathbf{m}_1 & \cdots & \mathbf{m}_N \\
        \mathbf{m}_1^\dagger & \mathbf{m}_0 & \cdots & \mathbf{m}_{N-1} \\
        \vdots & \vdots & \ddots & \vdots \\
        \mathbf{m}_N^\dagger & \mathbf{m}_{N-1}^\dagger & \cdots & \mathbf{m}_0
    \end{pmatrix}
    \succeq0 
    \label{eq:toeplitz}
\end{align}
is positive semidefinite for all $N$ \cite{Grenander_Toeplitz_1958}. Conversely, given any PSD $\mathcal{T}_N$, there exist (possibly infinitely many) spectral functions with those moments. Obtaining those requires the solution of a so-called truncated moment problem \cite{Akhiezer_moment_1965}.

Besides positivity and normalization, the Cayley-moment representation also preserves useful symmetry and boundedness properties. For the real-valued spectral matrices considered here,
since $f(-\omega)=f(\omega)^*$ on the real axis, a frequency-symmetric spectral component yields $\operatorname{Im}\mathbf{m}_k=0$, while a frequency-antisymmetric component yields $\operatorname{Re}\mathbf{m}_k=0$.
In addition, the moments inherit orbital-space symmetries: if $\mathbf{A}(\omega)$ is symmetric, then each $\mathbf{m}_k$ is also symmetric.
Under the Cayley transform in Eq.~\eqref{eq:cayley}, each lower-half-plane pole in Eq.~\eqref{eq:pole_rep} is mapped to a point inside the unit disk,
$\eta_\ell=f(\xi_\ell)$ with $|\eta_\ell|<1$. 
Since $|f(\omega)|=1$ on the real axis and $\mathbf{A}(\omega)$ is positive semidefinite, the moments remain bounded in operator norm by the total spectral weight encoded in $\mathbf{m}_0$.

A further advantage of the trigonometric moments is that a moment truncation can be readily converted to a compact pole representation.
Applying the residue theorem to the moment integral after the Cayley transform gives
\begin{align}
    \mathbf{m}_k=
    \int_{-\infty}^{\infty}d\omega\,
    \mathbf{A}(\omega)\,[f(\omega)]^k
    \simeq
    \sum_{\ell=1}^{M}\mathbf{A}_{\ell}\eta_\ell^k.
    \label{eq:mom_exp}
\end{align}
Recovering $\{\mathbf{A}_{\ell}, \eta_\ell\}$ from the moment sequence is a Prony-type approximation problem \cite{Prony_Prony_1795}.
We solve this problem using Estimation of Signal Parameters via Rotational Invariance Techniques (ESPRIT) \cite{Roy_ESPRIT_1989,Potts_ESPRIT_2013}, following Refs.~\cite{Zhang_minipole_2024_b, Zhang_minipole_2025} and \cite{Ying_pole_2022a,Ying_pole_2022b}.
The physical pole positions are then obtained by the inverse Cayley transform
\begin{align}
    \xi_\ell&=i\omega_p\frac{\eta_\ell+1}{\eta_\ell-1}.
    \label{eq:inv_pole}
\end{align}
The resulting set $\{\mathbf{A}_{\ell}, \xi_\ell\}$ yields a pole reconstruction of $\mathbf{G}^{\mathrm{R}}(z)$ (Eq.~\ref{eq:pole_rep}) that can be conveniently evaluated anywhere on the complex plane.
We emphasize that, while the construction via ESPRIT does in principle allow for a violation of positivity even if $\mathcal{T}_N$ is PSD, we have not observed such a violation in practice. Other methods for solving the truncated trigonometric moment problem exist but have not been explored here.

\begin{figure}[tbh]
    \centering
    \includegraphics[scale=1]{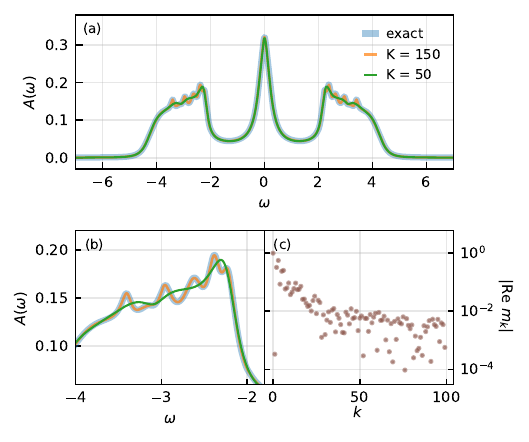}
    \caption{Systematic improvability of the moment representation. (a) Exact and reconstructed spectral functions for $K=50$ and $K=150$ moments agree closely over the full range. (b) A zoom into the fine oscillatory structure near $\omega\in[-4,-2]$ shows that $K=150$ resolves these features accurately, while $K=50$ smooths them out. (c) Magnitude of the real part of the moments, $|\operatorname{Re}(m_k)|$, versus $k$, showing the moment decay that governs how many moments are needed to resolve fine spectral detail.}
    \label{fig:mom_k}
\end{figure}
\begin{figure}[bth]
    \centering
    \includegraphics[scale=1]{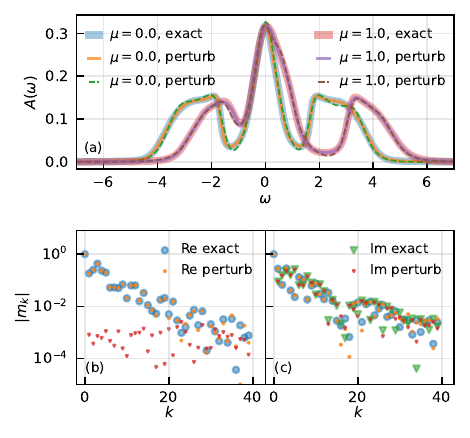}
    \caption{Robustness of the moment representation to noise. (a) Spectral functions at $U=4.0$ for $\mu=0.0$ and $\mu=1.0$, reconstructed from the exact 40-moment sequence and from sequences perturbed by independent random noise on the real and imaginary parts (solid: standard deviation $10^{-3}$; dashed: $10^{-2}$); reconstructions remain close to the exact spectra even under perturbation. (b, c) Real and imaginary parts of the exact and $10^{-3}$-perturbed moments for $\mu=0.0$ and $\mu=1.0$, respectively, showing that the perturbation is small relative to the unperturbed moment magnitudes.
    }
    \label{fig:mom_spec}
\end{figure}

For machine learning tasks that predict the Green's function, we represent the spectral matrix using a finite truncation of the moment sequence
$
\mathcal{M}_K=\{\mathbf{m}_0,\mathbf{m}_1,\ldots,\mathbf{m}_{K-1}\}.
$
Once $K$ is fixed, $\mathcal{M}_K$ is an ordered, fixed-dimensional collection of complex matrices, independent of the number of poles needed for reconstruction.
This representation also makes the physical constraints easy to impose. 
Equation~\eqref{eq:toeplitz} gives a direct way to enforce positivity in moment space. 
One may either project an unconstrained predicted moment sequence back onto the positive-semidefinite block-Toeplitz cone, analogous to the denoising procedure of Ref.~\onlinecite{Kemper_Denoise_2024}, or parameterize the moments so that this condition is satisfied by construction. 
Several by-construction parameterizations are possible, including fixed unit-circle discretizations with positive matrix weights \cite{DETTE_2010}, autocorrelation or spectral-factor coefficients \cite{Ephremidze_2009}, and Verblunsky or canonical-moment coefficients \cite{Simon_2005, DETTE_2010}.
For the machine learning prediction tasks in this work, we use the autocorrelation parameterization because it is compact, does not require choosing a fixed spectral grid, and gives a simple differentiable map from unconstrained neural-network outputs to normalized block-Toeplitz positive-semidefinite moments.

Specifically, instead of predicting the moments directly, the network outputs unconstrained coefficient matrices $\mathbf{B}_0,\ldots,\mathbf{B}_R$, which are mapped to moments through
\begin{align}
    \overline{\mathbf{m}}_k
    =
    \sum_{r=0}^{R-k}
    \mathbf{B}_{r+k}\mathbf{B}_{r}^{\dagger},
    \qquad k=0,\ldots,K-1 ,
    \label{eq:autocorr_mom}
\end{align}
where $R$ is a fixed hyperparameter satisfying $R\geq K$. 
In this work, we choose $R=\lfloor 1.5K\rfloor$ to give the autocorrelation parameterization additional flexibility. 
After normalization by $\overline{\mathbf{m}}_0^{-1/2}$, the resulting moments satisfy $\mathbf{m}_0=\mathbbm{1}$ and the block-Toeplitz positivity condition in Eq.~\eqref{eq:toeplitz} by construction. 
For real-symmetric spectral matrices, we additionally use the symmetrized form $(\overline{\mathbf{m}}_k+\overline{\mathbf{m}}_k^T)/2$.

In addition to this by-construction parameterization, we use projection onto the positive-semidefinite Toeplitz cone as a filter when subsequent operations produce unphysical moment sequences. 
Following Ref.~\onlinecite{Kemper_Denoise_2024}, this projection alternates between setting negative eigenvalues of the block-Toeplitz matrix to zero, restoring the Toeplitz structure by averaging along diagonals, and enforcing the normalization through the prescribed zeroth moment. 
This removes noncausal components before spectral reconstruction and yields a moment sequence consistent with a positive spectral measure.

\subsection{Moment Stability}
Figures~\ref{fig:mom_k} and \ref{fig:mom_spec} demonstrate that the moment representation is both systematically improvable and robust to small perturbations.
The spectra and moments used in these examples are generated from DMFT calculations on the Bethe lattice (see Section \ref{sec:single_dmft} for details).
Figure~\ref{fig:mom_k} shows an intermediate hybridization function from the DMFT self-consistency loop with damping, where the spectrum contains both broad structures and fine features before convergence.
The error level is set to $10^{-3}$ in the ESPRIT procedure used to recover the pole representation from the moment sequence in Eq.~\ref{eq:mom_exp}.
Increasing the number of retained moments from $K=50$ to $K=150$ improves the reconstruction of the finer features, while the low-order moments already capture the overall spectral envelope. 
For the applications considered here, $K$ is chosen empirically according to the spectral resolution required. 
$K=30$--$50$ is typically sufficient for smooth spectra, while intermediate DMFT spectra containing fine structures such as oscillatory features may require larger cutoffs.
Once the essential features are captured, the reconstruction is not sensitive to moderate further increases of $K$, and the main results remain stable over a broad range of moment cutoffs.

Figure~\ref{fig:mom_spec} uses converged DMFT spectra to test the stability of the moment representation under perturbations in moment space. 
Starting from the exact 40-moment sequence, we add random noise with standard deviations $10^{-3}$ and $10^{-2}$ (see panels~(b, c)) and reconstruct the corresponding spectra.
For the perturbed moment sequences, the ESPRIT reconstruction tolerance is set to $5\times10^{-3}$ for the $10^{-3}$ noise level and to $8\times10^{-2}$ for the $10^{-2}$ noise level.
These larger tolerances prevent the reconstruction from overfitting noise, since using an overly small tolerance for perturbed moments can introduce spurious poles and visible artifacts in the spectra.
As shown in panel~(a), the reconstructed spectral functions at $U=4.0$ remain close to the exact spectra for both $\mu=0.0$ and $\mu=1.0$ at the $10^{-3}$ noise level. 
At the larger noise level of $10^{-2}$, visible deviations appear in fine spectral features, but the overall spectral shape is still well preserved.
These results show that small perturbations in moment space lead to controlled changes in the reconstructed spectra, demonstrating that finite moment sequences provide a stable and robust representation of spectral functions.

\subsection{Model}

To benchmark the proposed moment representation, we consider multi-orbital Anderson impurity problems with a Slater-Kanamori interaction, both as standalone models and within DMFT, providing controlled examples of real-frequency Green's functions, hybridization functions, and self-energies with nontrivial spectral structure.
The Hamiltonian is
$\hat{H} = \hat{H}_\text{loc} + \hat{H}_\text{bath} + \hat{H}_\text{hyb}$, with
\begin{align}
    \hat{H}_\text{loc} &= \sum_{\nu_1\nu_2} 
    \varepsilon_{\nu_{1}\nu_{2}}\hat{d}_{\nu_{1}}^\dagger \hat{d}_{\nu_{2}} 
    + \hat{H}_\text{int},
    \\
    \hat{H}_\text{int}
    &=
    U \sum_i \hat n_{i\uparrow}\hat n_{i\downarrow}
    + \sum\limits_{i<j,\sigma\sigma^{\prime}}
    \left(U^{\prime}-J \delta_{\sigma \sigma^{\prime}}\right) 
    \hat{n}_{i \sigma} \hat{n}_{j \sigma^{\prime}}
    \nonumber\\
    &\quad
    - J \sum_{i\neq j}
    \hat d^\dagger_{i\uparrow}\hat d_{i\downarrow}
    \hat d^\dagger_{j\downarrow}\hat d_{j\uparrow}
    + J \sum_{i\neq j}
    \hat d^\dagger_{i\uparrow}\hat d^\dagger_{i\downarrow}
    \hat d_{j\downarrow}\hat d_{j\uparrow}, \label{eq:kanamori}
    \\
    \hat{H}_\text{bath} &= \sum_{b=1}^{N_b}
    \sum_{\kappa}\epsilon^b_{\kappa}\hat{c}_{b\kappa}^\dagger\hat{c}_{b\kappa},
    \\
    \hat{H}_\text{hyb} &= 
    \sum_{b=1}^{N_b}\sum_{\kappa \nu} 
    V^b_{\nu \kappa}\hat{d}_{\nu}^\dagger \hat{c}_{b\kappa} 
    + \mathrm{h.c.} \, .
\end{align}
Here, $U$ is the intra-orbital Hubbard interaction, $U'$ is the inter-orbital interaction, and $J$ is the Hund's coupling. For a rotationally invariant interaction, $U' = U-2J$. 
The operators $\hat{d}^{\dagger}_{i\sigma}$ and $\hat{d}_{i\sigma}$ create and annihilate fermions in impurity orbital $i$ with spin $\sigma\in \{\uparrow,\downarrow\}$. The index $b$ labels bath sites, while $\nu$ and $\kappa$ denote combined spin-orbital indices for the impurity and bath degrees of freedom.
The hybridization Hamiltonian $\hat{H}_\text{hyb}$ defines the coupling between the impurity and bath, with the corresponding hybridization function
\begin{align}
    \Delta_{\nu_{1}\nu_{2}}^R(\omega) = 
    \sum_{b=1}^{N_b} \sum_{\kappa} 
    \frac{V^{b}_{\nu_{1}\kappa} V^{b*}_{\nu_{2}\kappa}}
    {\omega + i 0^+ - \epsilon^{b}_{\kappa}}.
    \label{eq:hyb}
\end{align}
The retarded self-energy is obtained from the Dyson equation as
\begin{align}
    \mathbf{\Sigma}^R(\omega) =
    (\omega + \mu + i0^+)\mathbbm{1} - \boldsymbol{\varepsilon}
    - \mathbf{\Delta}^R(\omega) - \left[\mathbf{G}^R(\omega)\right]^{-1}.
\end{align}
The Green's function, self-energy, and hybridization function of this model are diagonal in spin space. 
Similar to the Green's function, the hybridization function admits a complex-pole representation and has an associated positive spectral density, although with a different normalization. 
The retarded self-energy has an analogous representation up to a static real contribution $\Sigma_\infty$ (the Hartree-Fock contribution), with its dynamical part described by a positive spectral density. 
We therefore also introduce moment representations for the hybridization function and the dynamical part of the self-energy, providing compact ways to encode the bath and correlation information.

\subsection{Real-frequency multi-orbital quantum impurity solvers}
Solving the DMFT equations in real frequency requires repeatedly obtaining the spectral function of an interacting quantum impurity model. This is a challenging quantum many-body problem. Imaginary-time methods, such as continuous-time quantum Monte Carlo \cite{Rubtsov05,Werner06,Werner06B,Gull08,Gull_CTQMC_2011}, rely on numerical analytic continuation and have therefore not been considered here.

For all the single-orbital DMFT benchmarks, we use the zero-temperature complex-time tensor-network impurity solver \cite{Ganahl_mps_2015,Bauernfeind_ftps_2017,Cao_tree_2021} introduced in Refs.~\cite{Cao_complextime_2024,Grundner_complextime_2024}. 
Its implementation within the real-frequency DMFT loop follows the procedure of Ref.~\onlinecite{Yu_complex_2026}. 
At each DMFT iteration, the solver performs time evolution along a complex-time contour and extracts real-frequency spectra through an exponential-fitting procedure, enabling us to generate high-quality real-frequency data efficiently for the single-orbital benchmarks.

For the two-orbital matrix-valued benchmark, we use a steady-state noncrossing approximation (NCA) impurity solver \cite{eckstein_nonequilibrium_2010,cohen_greens_2014,erpenbeck_revealing_2021,erpenbeck_resolving_2021,Zemach_NCA_2024}.
A multitude of highly accurate multi-orbital quantum impurity solvers exist, such as the  inchworm quantum Monte Carlo method \cite{cohen_taming_2015,eidelstein_multiorbital_2020,erpenbeck_quantum_2023,erpenbeck_steady-state_2024} which obtains numerically exact results at non-zero temperature.
The NCA solver used in this work is approximate, but provides a rapid, practical, and efficient way to generate large datasets of physically reasonable matrix-valued spectral functions with off-diagonal orbital structure.
In this setup, the hybridization scale is chosen relatively large so that the propagators decay within the accessible simulation time and the steady state is reached reliably. See Appendix~\ref{app:two_orb_method} for more details.

We emphasize that the moment representation and the machine-learning formulation are independent of the  impurity solver and can therefore be applied to real-frequency Green's functions obtained from any real-axis impurity solver or any many-body method.

\subsection{Neural network}

\begin{figure}[bt]
    \centering
    \includegraphics[width=1\linewidth]{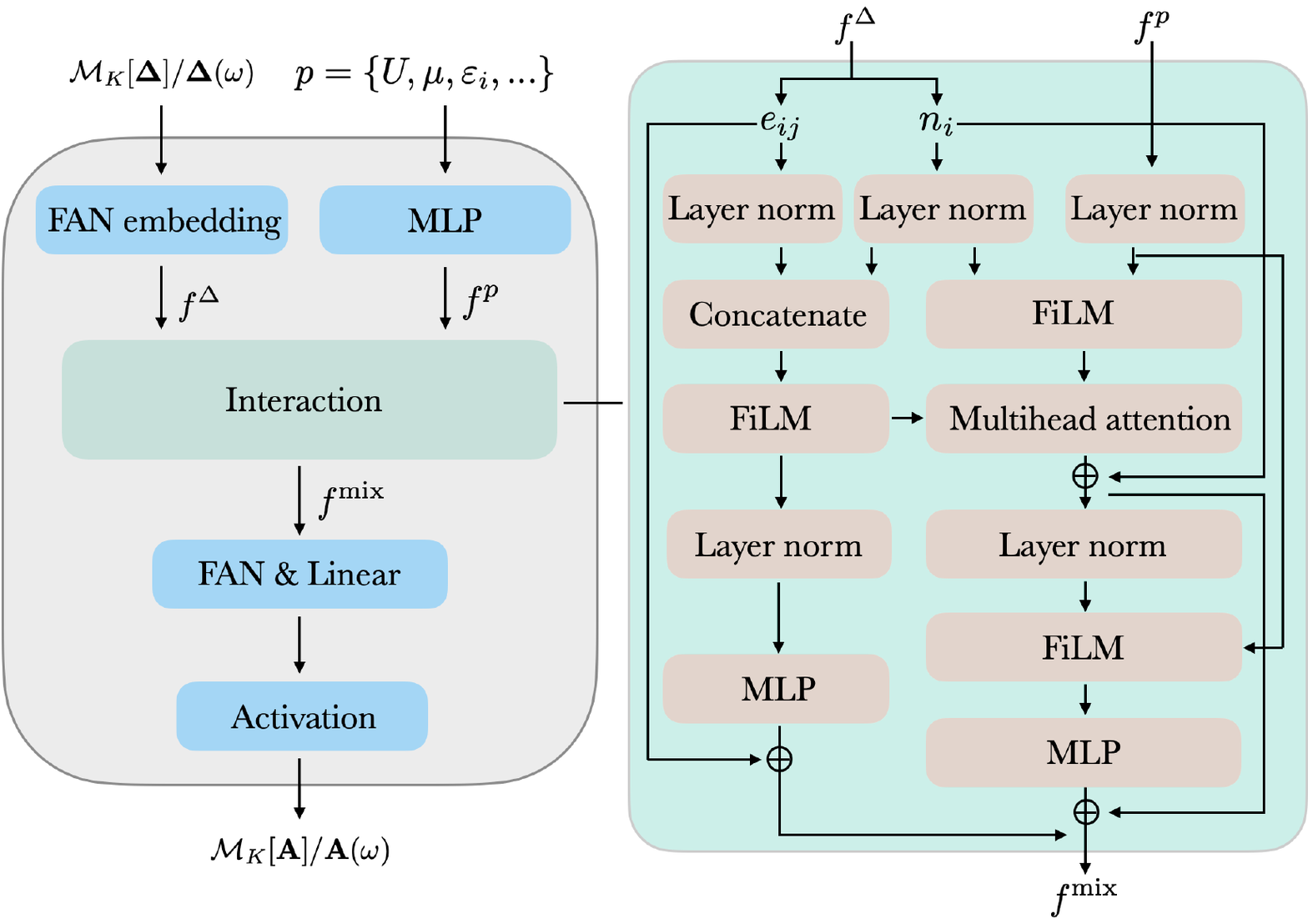}
    \caption{Schematic of the neural network architecture. Left: the three modules of the network -- input embedding (FAN embedding of the hybridization moments $\mathcal{M}_K[\Delta]$, MLP embedding of the physical parameters $p$), the FiLM-conditioned graph interaction block (green), and the output projection (FAN \& Linear followed by an activation, yielding either the predicted moment sequence $\mathcal{M}_K[A]$ or direct frequency-domain spectral values). The interaction block is repeated $L$ times, with the output $f^\text{mix}$ of one layer used as the input $f^{\Delta}$ of the next. Right: detailed structure of the interaction block, showing the edge ($e_{ij}$) and node ($n_i$) update pathways under FiLM conditioning on the physical-parameter embedding $f^p$. For machine-learning terminology (FAN, MLP, FiLM), see main text.
    }
    \label{fig:network}
\end{figure}

We choose a neural network architecture that reflects the structure of the impurity problem.
The hybridization-to-spectrum map is nonlinear and depends on the full moment sequence and physical parameters, so we use an attention-based architecture rather than a simple feed-forward mapping of flattened features to mix moment and parameter information more flexibly.
In a multi-orbital system, the Green's function, hybridization function, and spectral function are matrix-valued objects: diagonal elements describe orbital-resolved local spectra, while off-diagonal elements encode orbital mixing.
We therefore represent each matrix-valued moment sequence as a graph over spin-orbital degrees of freedom, with diagonal terms treated as node features and off-diagonal terms treated as edge features.

Within this graph representation, we map the input moment sequence $\{\mathbf{m}_k(\boldsymbol{\Delta})\}$ and physical parameters $p=\{U,\mu,\varepsilon_i,\ldots\}$ to the target moment sequence $\{\mathbf{m}_k(\mathbf{A})\}$ using an attention-based graph-network architecture inspired by graph neural networks \cite{gilmer_graph_2017,battaglia_graph_2018,velickovic_graph_2018}, together with Feature-wise Linear Modulation (FiLM) conditioning \cite{perez_film_2018}.
The attention mechanism learns correlations among node and edge features, while FiLM conditioning incorporates the physical parameters as global conditions in the graph updates.
For comparison, we also allow the input and output features to be direct frequency-domain data points $\{\boldsymbol{\Delta}(\omega)\}$ and $\{\mathbf{A}(\omega)\}$ instead of moments.

Figure~\ref{fig:network} schematically illustrates the neural-network architecture.
The left panel shows the overall data flow, while the right panel zooms in on the conditioned graph-interaction block.
We describe the network in three modules: before the interaction block, the input embedding converts the hybridization data and physical parameters into latent features; in the interaction block, node and edge information is mixed under physical parameter conditioning; after the interaction block, the output projection maps the resulting representation to either spectral moments or direct frequency-domain spectral data.
Details for the three modules are provided below.

\paragraph*{Input embedding}

The physical parameters $p$ are embedded by a multi-layer perceptron (MLP)
\begin{align}
    &p^{(l)} = \text{Swish}(W^{(l)} p^{(l-1)} + b^{(l)}), 
    \\
    &f^{p} = (W^{(L)} p^{(L-1)} + b_L),
\end{align}
with $p^{(0)} = p$, $l=1,...L-1$, where $L$ is the number of layers and $W$ and $b$ are trainable parameters.
For input hybridization, the real and imaginary parts are concatenated and embedded by $L$ Fourier Analysis Network (FAN) layers
\begin{align}
&q^{(l)} = W_q^{(l)} x^{(l)} + b_q^{(l)}, 
\quad 
g^{(l)} = W_g^{(l)} x^{(l)} + b_g^{(l)} ,
\\
&f^{\Delta, (l)} = \left[\sin(q^{(l)}), \cos(q^{(l)}), {\rm GELU}(g^{(l)}) \right],
\end{align}
with $x^{(0)} = \{\mathrm{Re}[\mathcal{M}_K (\mathbf{\Delta})], \mathrm{Im}[\mathcal{M}_K (\mathbf{\Delta})]\}$ or $x^{(0)} = \{\mathbf{\Delta}(\omega)\}$. 
The embedded quantities $f^{p}$ and $f^{\Delta}$ serve as the encoded physical-parameter and hybridization representations.
The diagonal components of $f^{\Delta}$ are used to initialize the node features $n_i^{(0)}$, while the off-diagonal components are used to initialize the edge features $e_{ij}^{(0)}$.

\paragraph*{Interaction}
The interaction block mixes node and edge features conditioned on the physical parameters using FiLM conditioning
\begin{align}
    \mathrm{FiLM}(x;c) =
    (1 + \left(W_{\gamma}c+b_{\gamma}\right))\odot x
    +
    \left(W_{\beta}c+b_{\beta}\right) ,
\end{align}
where $\odot$ denotes elementwise multiplication.
The input node, edge, and parameter features first go through layer normalization (LN), and the network then applies graph update blocks. In block $l$, the edge update is computed from the node and edge features and then conditioned on the physical parameters
\begin{align}
    &r_{ij}^{(l)} =
    \left[ \tilde{e}_{ij}^{(l)},
    \tilde{n}_i^{(l)}, \tilde{n}_j^{(l)} \right],
\quad
    &\bar r_{ij}^{(l)} = \mathrm{FiLM}(r_{ij}^{(l)}; \tilde{f}^p),
    \label{eq:edge_concat}
\end{align}
with $\tilde{X} = \mathrm{LN}(X)$ and $i$, $j$ denoting node labels.
The output edge features are computed with a residual update
\begin{align}
    e_{ij}^{(l+1)} = e_{ij}^{(l)} +
    \mathrm{MLP} \left( \mathrm{LN}(\bar r_{ij}^{(l)}) \right).
\end{align}
The node updates are performed by edge-conditioned multi-head attention after FiLM conditioning
\begin{align}
    \bar n_i^{(l)} =
    \mathrm{FiLM} ( \tilde{n}_i^{(l)}; \tilde{f}^{p} ).
\end{align}
For head \(h\),
\begin{align}
    q_i^h = W_Q^h \bar n_i^{(l)},
    \quad
    k_i^h = W_K^h \bar n_i^{(l)},
    \quad
    v_i^h = W_V^h \bar n_i^{(l)},
\end{align}
and the attention logits include an edge-dependent bias
\begin{align}
    &\alpha_{ij}^{h} =
    \mathrm{softmax}_{j}
    \left[ \frac{q_i^h\cdot k_j^h}{\sqrt{d_h}}
    + b_{ij}^{h} \right], \quad 
    &b_{ij}^h = W_b^h \bar{r}_{ij}^{{(l)}}.
\end{align}
The attention output of layer $l$ is given by
\begin{align}
    u_i^{(l)} = W_O \left[ \bigoplus_h \sum_j \alpha_{ij}^{h} (v_j^h + W_E^h \bar{r}_{ij}^{{(l)}}) \right],
\end{align}
where $\bigoplus_h$ denotes concatenation over heads.
The output node features are computed as
\begin{align}
    &\hat n_i^{{(l)}} = n_i^{{(l)}}+u_i^{{(l)}}, 
    \\
    &n_i^{(l+1)} = \hat n_i^{{(l)}} + \mathrm{MLP}
    \left[ \mathrm{FiLM} \left(
    \mathrm{LN}(\hat n_i^{{(l)}}); \tilde{f}^p \right) \right].
\end{align}
The interaction block is repeated $L$ times, with the output features of each layer used as the input features of the next layer.

\paragraph*{Output projection}
After the final graph block, the node and edge features are collected into a mixed representation $y^\text{mix}$ and passed through another FAN layer followed by a linear layer
\begin{align}
    \tilde{y}
    =
    W_{\mathrm{out}}
    \left( \mathrm{FAN}(y^\text{mix}) \right)
    +
    b_{\mathrm{out}} .
\end{align}
The interpretation of $\tilde{y}$ depends on the output representation. 
For moment prediction, $\tilde{y}$ is reshaped into unconstrained autocorrelation coefficients $\{\mathbf{B}_r\}$, which are mapped to normalized positive-semidefinite moments by the autocorrelation layer defined in Eq.~\eqref{eq:autocorr_mom}. 
The moment output is then
\begin{align}
    y^\text{out} =\{\mathrm{Re}[\mathcal{M}_k (\mathbf{A})], \mathrm{Im}[\mathcal{M}_k (\mathbf{A})]\}.
\end{align}
For direct frequency-domain prediction, we instead apply a softplus activation to enforce non-negativity of the predicted spectral function,
\begin{align}
    y^\text{out}
    =
    \{\mathbf{A}(\omega)\}
    =
    \operatorname{softplus}(\tilde{y}).
    \label{eq:softplus_out}
\end{align}

Given the network output $y^\text{out}$ defined above, we train the model by matching it to the corresponding target output, either in moment space or in the direct frequency domain. For each data point $x$, we compute a data loss by summing the squared Frobenius norms of the differences between the predicted and target output matrices. 
The total data loss is then obtained by averaging over all $N$ data points. 
To reduce overfitting, we add an $L_2$ regularization term with coefficient $\lambda=10^{-8}$ over the trainable parameters $\theta$,
\begin{align}
    \mathcal{L}
    &=
    \frac{1}{N}\sum_{x \in \mathcal{D}} \mathcal{L}_{\mathrm{data}}(x)
    + \lambda \|\theta\|_2^2, 
    \\
    \mathcal{L}_{\mathrm{data}}(x)
    &=
    \sum_k
    \left\|
    y_{k,\mathrm{pred}}^{\mathrm{out}}(x)
    -
    y_{k,\mathrm{label}}^{\mathrm{out}}(x)
    \right\|_F^2 .
\end{align}
We also apply dropout with a rate of $0.02$ as an additional regularization mechanism.

\section{Results}

We assess the moment representation through three progressively more challenging benchmarks. 
The first benchmark consists of single-orbital DMFT calculations on the Bethe lattice for both half-filled and doped systems. 
At half filling, we compare moment-space learning with direct frequency-domain learning to test whether the compact moment representation retains the accuracy of a direct frequency-domain target.
Away from half filling, we examine whether the predicted spectra reproduce physical observables such as the density. 
The second benchmark uses antiferromagnetic single-orbital DMFT on the cubic lattice, where the learned impurity map is embedded in a real-frequency self-consistency loop and the self-energy is reconstructed through the Dyson equation. 
This tests the representation in a symmetry-broken phase and probes its robustness under self-energy feedback. 
The third benchmark applies the same framework to a two-orbital impurity model to test whether the matrix-valued moment representation captures orbital-resolved and off-diagonal spectral structure. 
Details of the neural-network architecture and training settings used in these benchmarks are provided in Appendix~\ref{app:network}.

\subsection{Single-orbital DMFT on the Bethe lattice} \label{sec:single_dmft}

We first benchmark the moment representation using single-orbital zero-temperature real-frequency DMFT on the Bethe lattice in the infinite-coordination limit. 
The DMFT loop is closed by the Bethe-lattice self-consistency condition
$\Delta^R(\omega)=\frac{D^2}{4}G^R(\omega)$,
which relates the hybridization function to the impurity Green's function. 
Here, $D=2t$ is the half bandwidth of the noninteracting semicircular density of states
$A_0(\omega)=\frac{\sqrt{4t^{2}-\omega^{2}}}{2\pi t^{2}}$, and $t$ is the hopping amplitude that sets the energy unit.
\begin{figure}[tb]
    \centering
    \includegraphics[scale=1]{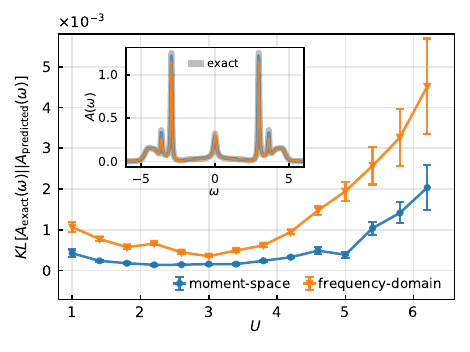}
    \caption{Comparison of moment-space and frequency-domain learning for the half-filled Bethe-lattice benchmark. KL divergence between predicted and exact spectral functions, binned in $U$ with bin width 0.4 and error bars showing the spread within each bin; the moment-space representation matches or modestly outperforms the higher-dimensional frequency-domain representation across the full range of $U$. Inset: exact and predicted spectral functions (moment- and frequency-domain, colors as in the main panel) at $U=6.25$, third DMFT iteration, illustrating the sharp features responsible for the larger errors at strong coupling.
    }
    \label{fig:kl_1orb}
\end{figure}

At half filling, we compare the moment representation with the direct real-frequency-domain representation.  The DMFT runs used to generate the data use a mixing factor of 0.5.
Details of the dataset construction are given in the Appendix~\ref{app:sdmft}. 
All hybridization functions $\Delta^R(\omega)$ and spectral functions $A(\omega)$ are stored on an equidistant frequency grid with 2001 points for $\omega\in[-10,10]$ and converted to moments using $\omega_p=2$ (half bandwidth) up to $K=150$.

Figure~\ref{fig:kl_1orb} compares the accuracy of moment-space and frequency-domain learning for the half-filled Bethe-lattice benchmark. 
The moment model is trained with $K=150$ input and output moments, since the hybridization functions require this number of moments for reliable reconstruction (see Figure~\ref{fig:mom_k}). 
The predicted spectral functions are reconstructed using the first 100 moments with an ESPRIT tolerance of $5\times10^{-3}$, which is sufficient to capture their smooth structure while avoiding small unphysical spikes from errors in higher-order moments.
The error is quantified by the Kullback--Leibler (KL) divergence,
$\mathrm{KL}[A_1\|A_2]=\int d\omega\,A_1(\omega)\ln\!\left[A_1(\omega)/A_2(\omega)\right]$,
between the exact and predicted spectral functions and is averaged over interaction-strength bins of width $\Delta U=0.4$.
Both representations achieve small errors at weak and intermediate interaction strengths, and their performance remains broadly comparable as $U$ increases, with the moment representation showing a modest advantage.
The larger errors at strong coupling mainly arise from the first few iterations of the large-$U$ simulations, where the spectral functions contain sharp peaks, as shown in the inset.

\begin{figure}[tb]
    \centering
    \includegraphics[scale=1]{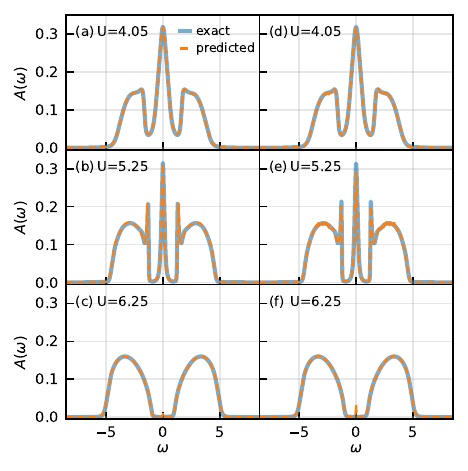}
    \caption{Converged spectral functions from the undamped, self-consistent ML-DMFT loop at representative interaction strengths $U=4.05$, $5.25$, and $6.25$, using moment-space (a--c) and frequency-domain (d--f) predictions. Both representations converge to physically reasonable spectra, but the moment-space representation shows modestly better agreement with the exact spectra in resolving the sharp low-energy and Hubbard-band features near the Mott transition (compare panels (b) and (e)).
    }
    \label{fig:iterative_1orb}
\end{figure}

Despite these larger early-iteration errors, the ML-DMFT loop remains stable for both representations. 
Figure~\ref{fig:iterative_1orb} shows the converged spectra obtained after iterative ML-DMFT prediction for representative interaction strengths. 
In these evaluations, we use no damping, so each updated hybridization function is generated directly from the model-predicted Green's function without reference to earlier iterations.
This makes the hybridization trajectories differ more strongly from the mixed DMFT histories used for training and provides a stricter test of stability. 
For the representative interaction strengths shown in Figure~\ref{fig:iterative_1orb}, the ML-DMFT loop is run for 10, 10, and 20 iterations depending on the interaction strength.
The converged spectral functions are constructed from 30 moments using an ESPRIT tolerance of $2\times10^{-3}$, since the converged DMFT spectra are smooth and do not contain fine features.
Both the moment-space and frequency-domain models converge to reasonable spectral functions, indicating that the prediction errors in Fig.~\ref{fig:kl_1orb} do not prevent self-consistency. 
Nevertheless, the moment representation gives slightly better agreement with the exact spectra, especially in resolving the sharp low-energy and Hubbard-band features near the phase transition.

\begin{figure}[tb]
    \centering
    \includegraphics[scale=1]{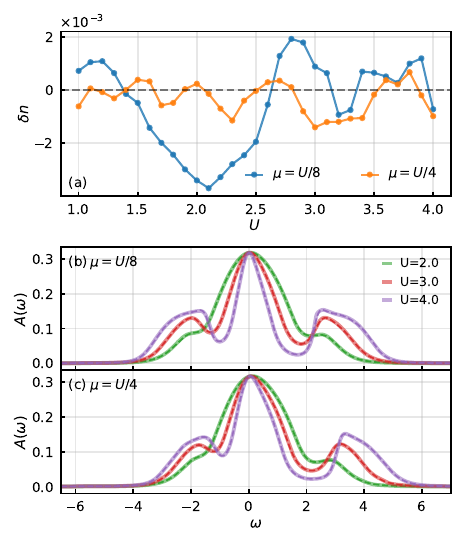}
    \caption{Density and spectral accuracy away from half filling. (a) Density error $\delta n = n_\text{pred}-n_\text{label}$ vs. $U$ for $\mu=U/8$ (held out from training) and $\mu=U/4$ (included in training); both remain on the order of $10^{-3}$, comparable to the DMFT convergence criterion used to generate the data. (b,c) Exact (solid) and predicted (dashed) converged spectral functions at representative $U$ values for $\mu=U/8$ (b) and $\mu=U/4$ (c), showing that the overall spectral shape remains accurately captured away from particle-hole symmetry.}
    \label{fig:density}
\end{figure}

As a further test, we examine whether the moment representation can accurately reproduce the density, which in the past has been difficult to obtain from Matsubara frequency predictions
\cite{Dong_equivGF_2024,Valenti_NNembedding_2026}.
For the doped Bethe-lattice calculations, we use the same frequency grid and moment parameters as in the half-filled benchmark.
The converged spectra are constructed from 30 moments using an ESPRIT tolerance of $2\times10^{-3}$.
Further details of the dataset construction are provided in Appendix~\ref{app:sdmft}.

Figure~\ref{fig:density} shows the converged ML-DMFT results away from half filling.
As in the half-filled benchmark, we run 10 iterations of the ML-DMFT loop with a mixing factor of $1.0$, so the hybridization trajectory is generated directly by the trained model and differs from the mixed DMFT histories used for training.
Panel~(a) shows the density error $\delta n = n_\text{pred} - n_\text{label}$ for $\mu=U/8$ and $\mu=U/4$ across the interaction range.
The training branch $\mu=U/4$ gives smaller errors, while the held-out branch $\mu=U/8$ shows somewhat larger deviations.
The held-out errors nevertheless remain small, staying on the order of $10^{-3}$, comparable to the convergence criterion used when generating the DMFT data. 
This indicates that the reconstructed spectra accurately preserve the occupied spectral weight.
Panels~(b) and~(c) compare exact and predicted converged spectra at representative $U$ values for the two dopings.
Small discrepancies appear in some high-frequency fine features, but the overall spectral shape is well captured, showing that the moment-based model remains accurate when particle-hole symmetry is broken.

\subsection{Antiferromagnetic DMFT on the cubic lattice}

\begin{figure}[tb]
    \centering
    \includegraphics[scale=1]{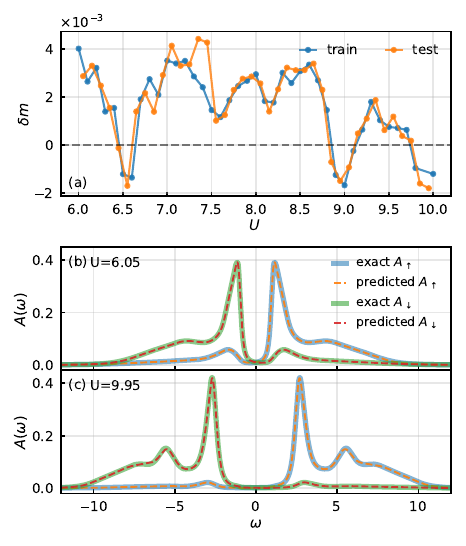}
    \caption{Antiferromagnetic order and spin-resolved spectral accuracy on the cubic lattice. (a) Staggered-magnetization error $\delta m = m_\text{pred}-m_\text{label}$ vs. $U$ for training and validation interaction strengths, remaining on the order of $10^{-3}$ throughout. (b,c) Exact (solid) and predicted (dashed) spin-resolved spectral functions $A_\uparrow(\omega)$ and $A_\downarrow(\omega)$ at representative interaction strengths $U=6.05$ (b) and $U=9.95$ (c), showing that the antiferromagnetic mirror symmetry between spin channels is accurately reproduced.}
    \label{fig:magnet}
\end{figure}

To further benchmark the moment representation beyond the Bethe lattice examples, we consider a symmetry-broken setting: zero-temperature antiferromagnetic DMFT for the half-filled single-orbital Hubbard model on a three-dimensional cubic bipartite lattice. 
The calculations are carried out using a two-sublattice DMFT loop.
We use nearest-neighbor hopping $t=1$, which sets the energy unit, and set the chemical potential to $\mu=U/2$. Details of the simulations and dataset construction are given in Appendix \ref{app:afm_dmft}.
For this benchmark, $\boldsymbol{\Delta}^R(\omega)$ and $\mathbf{A}(\omega)$ are stored on a 5001-point uniform grid over $\omega\in[-20,20]$ and converted to moments up to $K=50$ using $\omega_p=6$ (half bandwidth).

Unlike the Bethe-lattice benchmarks, the cubic-lattice self-consistency requires computing the self-energy from the predicted $\mathbf{G}^R(\omega)$ and $\boldsymbol{\Delta}^R(\omega)$ through the Dyson equation, as described in Appendix~\ref{app:afm_dmft}. 
Because this step involves inverting $\mathbf{G}^R(\omega)$, small prediction errors can be strongly amplified when $\mathbf{G}^R(\omega)$ is close to zero, leading to noncausal or oscillatory features in $\boldsymbol{\Sigma}^R(\omega)$.
We therefore regularize the self-energy in moment space. 
After computing $\boldsymbol{\Sigma}^R(\omega)$ from the Dyson equation, we retain the first 20 moments of its dynamical part, project this truncated moment sequence onto the positive-semidefinite Toeplitz cone as described in Sec.~\ref{sec:mom}, and reconstruct the filtered self-energy. 
The truncation suppresses high-order noisy components, while the Toeplitz projection enforces consistency with a positive spectral measure. 
As a comparison, a direct real-frequency causal filter based on clipping the noncausal part of $\operatorname{Im}\boldsymbol{\Sigma}^R(\omega)$ and recomputing $\operatorname{Re}\boldsymbol{\Sigma}^R(\omega)$ through the Kramers--Kronig relation gives similar results but slightly larger errors. 
For the ESPRIT reconstructions of both $\mathbf{G}^R(\omega)$ and the dynamical part of $\boldsymbol{\Sigma}^R(\omega)$, we use a tolerance of $5\times10^{-3}$.

Figure~\ref{fig:magnet} shows the converged antiferromagnetic DMFT results.
The upper panel shows the magnetization error $\delta m=m_{\mathrm{pred}}-m_{\mathrm{label}}$ for both training and validation interaction strengths. 
The errors remain on the order of $10^{-3}$, indicating that the learned impurity map consistently reproduces the staggered magnetization across the sampled $U$ range.
The lower panels compare representative spin-resolved spectral functions at the smallest and largest test interaction strengths, $U=6.05$ and $U=9.95$, showing close agreement between the predicted and reference spectra for the spin components related by antiferromagnetic mirror symmetry. 
Although some high-frequency fine structures show minor deviations, the overall spin-resolved spectral profiles are reproduced accurately.
These results show that the moment-based impurity map remains stable in a symmetry-broken DMFT loop and accurately captures both the antiferromagnetic order parameter and the underlying spin-resolved spectral structure.

\subsection{Multi-orbital steady state impurity model}

As a proof of concept for matrix-valued spectral functions, we test the moment representation using a two-orbital steady-state impurity model solved with the NCA solver.
The input hybridization functions are generated from random positive-definite matrix-valued pole expansions, producing spectra with both diagonal and off-diagonal orbital components. 
The steady-state NCA calculations use the two-orbital Kanamori interaction in Eq.~\ref{eq:kanamori}, with impurity parameters sampled over a range of interaction strengths and dopings near half filling. 
Details of the hybridization construction, parameter sampling, and NCA setup are provided in Appendix~\ref{app:two_orb_data}.

\begin{figure}[tb]
    \centering
    \includegraphics[scale=1]{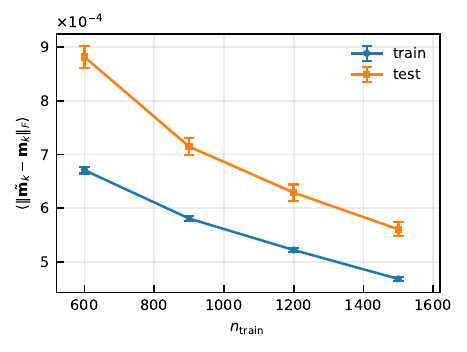}
    \caption{Learning curve for the two-orbital model. Average Frobenius-norm error between predicted and exact moment matrices, $\langle\|\tilde m_k - m_k\|_F\rangle$, for the training and test sets as a function of the number of training samples $n_\text{train}$. Both errors decrease monotonically over the tested range, with the test error remaining consistently above the training error, indicating that the model continues to benefit from additional training data without saturating within this range.
    }
    \label{fig:error_2orb}
\end{figure}
\begin{figure}[bt]
    \centering
    \includegraphics[scale=1]{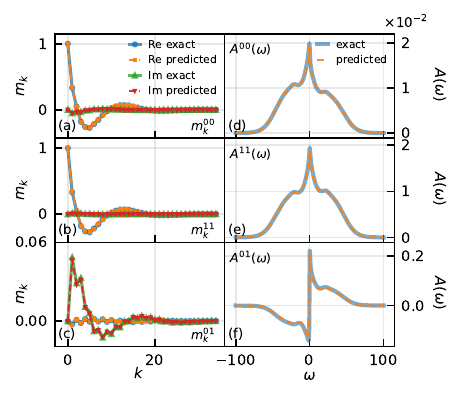}
    \caption{Worst-case test-sample reconstruction for the two-orbital model (1500 training samples), at the test point with the largest average moment error ($U\approx3.40$, $\mu\approx4.37$). (a--c) Real and imaginary parts of the exact and predicted moments for the diagonal components $m_k^{00}$, $m_k^{11}$ and the independent off-diagonal component $m_k^{01}$ (note the different vertical scale in panel (c), reflecting the much smaller magnitude of the off-diagonal moments). (d--f) Corresponding exact and predicted spectral functions $A^{00}(\omega)$, $A^{11}(\omega)$, and the sign-changing off-diagonal component $A^{01}(\omega)$. Even for this worst-case example, the predicted moments and spectra closely track the exact results, including the off-diagonal orbital-mixing structure.
    }
    \label{fig:spec_2orb}
\end{figure}

After generating the matrix-valued spectra, we convert each spectrum into a moment sequence with $K=50$ using $\omega_p=10$.
Since the off-diagonal moment components are typically smaller than the diagonal ones but are essential for describing orbital mixing, we rescale them during training by an integer prefactor $q=\lfloor 1/m_{\max}\rfloor$, where $m_{\max}$ is the largest absolute value among the real and imaginary components throughout the entire dataset.
This rescaling is applied only when computing the loss for the output prediction, not to the input hybridization moments.
This brings the largest rescaled off-diagonal component close to unity and improves the numerical balance between diagonal and off-diagonal features. 

Using these processed moment sequences as learning targets, we examine how the prediction accuracy depends on the training-set size.
For a dataset $\mathcal{S}$, we measure the prediction error by averaging the Frobenius-norm error over both data points and moment index
\begin{align}
    \langle \|\tilde{\mathbf{m}}_{k} - \mathbf{m}_{k}\|_F \rangle
    =
    \frac{1}{N_{\mathcal{S}}K}
    \sum_{i=1}^{N_{\mathcal{S}}}
    \sum_{k=0}^{K-1}
    \left\|
    {\mathbf{m}}_{k,\text{pred}}^{(i)}
    -
    \mathbf{m}_{k,\text{exact}}^{(i)}
    \right\|_F .
\end{align}
Figure~\ref{fig:error_2orb} shows the average Frobenius norm error of the predicted moment matrices for both the training and test sets. 
As $n_{\mathrm{train}}$ is increased from 600 to 1500, both errors decrease systematically, suggesting that the model learns meaningful structure from the NCA data and benefits from additional training samples. 
We therefore use the model trained with 1500 samples for the representative reconstruction shown below.

Figure~\ref{fig:spec_2orb} shows the test sample with the largest average moment error for the model trained on 1500 samples. 
Even for this worst-case example, the predicted moments closely follow the real and imaginary parts of the exact diagonal components, $m_k^{00}$ and $m_k^{11}$, as well as the independent off-diagonal component $m_k^{01}$. 
Because the spectral and moment matrices are symmetric, the moment matrix inherits this symmetry, so $m_k^{10}=m_k^{01}$ and the off-diagonal entries are fully specified by $m_k^{01}$ alone.
The reconstructed spectra reproduce the diagonal spectral functions $A^{00}(\omega)$ and $A^{11}(\omega)$, together with the sign-changing off-diagonal component $A^{01}(\omega)$.
Some fine features are not perfectly resolved, but the overall diagonal and off-diagonal spectral shapes are correctly reconstructed.
The accurate reconstruction of both diagonal and off-diagonal spectral components demonstrates that the moment representation can capture the full matrix-valued structure of Green's functions. 

\section{Conclusions and Discussion} \label{sec:conclusion}

In this manuscript, we introduced a moment-based representation for spectral functions designed for machine-learning applications. 
The representation is built from Cayley-mapped trigonometric moments with the Jacobian included, so that the zeroth moment retains the physical spectral-weight normalization and the moment sequence remains tied to a positive matrix-valued spectral measure. 
By targeting real-frequency spectra directly, this approach avoids the ill-posed analytic-continuation step required in Matsubara-frequency-based frameworks, which is particularly important for observables such as the density that are sensitive to high-frequency errors. 
Compared with a direct real-frequency representation on a dense frequency grid, the finite moment sequence provides a more compact and efficient description while preserving physical constraints such as normalization and positivity. 
The representation also avoids the ambiguity and variable size of direct pole coordinates: once the cutoff $K$ is chosen, the moment sequence has a fixed dimension and can still be converted back to a real-frequency spectrum through an ESPRIT-based pole reconstruction.

We demonstrated this representation using a graph-neural-network-inspired architecture and impurity/DMFT benchmarks. 
Across the single-orbital and two-orbital benchmarks studied here, the moment-based framework accurately reproduces spectral features, preserves the density away from half filling, and extends naturally to matrix-valued spectra with orbital mixing. 
In the antiferromagnetic cubic-lattice benchmark, projecting the dynamical self-energy moments onto the positive-semidefinite Toeplitz cone suppresses noncausal artifacts from Dyson-equation inversion and enables stable real-frequency self-consistency beyond the Bethe-lattice setting. 
Together, these results show that finite moment sequences provide compact, stable, and physically constrained learning targets for real-frequency spectral data.

Several aspects remain to be explored. 
The benchmarks presented here involve at most two orbitals, while realistic multi-orbital electronic-structure problems may involve tens to hundreds of orbitals. 
In such settings, the compactness of the moment representation is expected to be especially useful: although the matrix structure still scales with the number of orbitals, the frequency dependence is represented by a finite moment sequence rather than dense frequency-grid data. 
The practical scaling of the required moment cutoff $K$, the ESPRIT-based pole reconstruction, and the network size with orbital number remains to be tested. 
Nevertheless, related matrix-valued pole-reconstruction techniques have already been successfully applied to analytic continuation of Green's functions and self-energies in realistic materials calculations \cite{Zhang_minipole_2024_b}, suggesting that the reconstruction step itself should not pose a fundamental limitation.
We also note that, while the autocorrelation construction guarantees a positive-semidefinite block-Toeplitz moment sequence, the subsequent ESPRIT-based pole reconstruction does not strictly preserve positivity by construction. 
In practice, we did not observe positivity violations, and other numerical methods exist that guarantee positive spectral functions  \cite{Mead_maxent_1984,Zhang_minipole_2024_b}.

Looking forward, the moment representation may be combined with other machine-learning architectures, extended to other real-frequency response functions such as susceptibilities, and used in steady state transport settings where compact real-frequency representations are similarly advantageous.

\begin{acknowledgments}
X.D. would like to thank Fenglin Deng for discussions.
X.D. is funded by the National Natural Science Foundation of China under grant No. 12504289. EG
acknowledges funding from the European Research Council
(ERC) under Advanced Grant No. 101142136 (Quantum Algorithms).
G.C. acknowledges support from the Israel Science Foundation (Grant No. 2902/21), the PAZY Foundation (Grant No. 318/78) and the MOST NSF-BSF (Grant No. 2023720).
L. Zhang was supported by the National Science Foundation under Grant No. NSF QIS 2310182.
\end{acknowledgments}

\appendix

\section{Network and training details} \label{app:network}

For the single-orbital benchmarks, the moment-space and frequency-domain models use the same network widths and embedding structure within each benchmark. 
The hybridization embedding consists of two FAN layers with widths $[128,128]$ for the half-filled and antiferromagnetic benchmarks, and $[256,128]$ for the doped benchmark. 
The physical-parameter embedding uses two layers with widths $48$ and $128$. 
Since the single-orbital problem contains only one diagonal component, no edge embedding is used. 
The graph representation has width $128$, with two FiLM-conditioned graph blocks and four attention heads in each block. 
The output projection uses one hidden layer, with width $128$ for the half-filled and antiferromagnetic benchmarks and width $256$ for the doped benchmark.

For the two-orbital NCA benchmark, we use the same physical-parameter embedding width, graph width, number of graph blocks, number of attention heads, and output hidden width. 
The main difference is that the matrix-valued input contains off-diagonal components, so both node and edge embeddings are used. 
The node and edge embeddings each consist of two FAN layers with widths 64 and 128.

All models are trained for 25000 epochs. 
For the single-orbital benchmarks, the initial learning rate is $5\times10^{-3}$, while for the two-orbital NCA benchmark it is reduced to $5\times10^{-4}$. 
In all cases, the learning rate is decayed every 100 steps by a factor of $0.98$, with a minimum learning rate of $10^{-6}$.

\section{Single-orbital Bethe lattice DMFT data details} \label{app:sdmft}

For the half-filled benchmark, the data are generated for $U\in[1.0,6.4]$ in steps of $0.05$. 
The training set initially contains $U=1.0,1.1,\ldots$, while the test set contains the interleaved values $U=1.05,1.15,\ldots$. 
For $U>5.0$, every other interleaved value such as $U=5.15,5.35,\ldots$ is additionally included in the training set to better sample the more challenging large-$U$ regime.
Each DMFT run uses a mixing factor of $0.5$ and is carried out for 10 iterations for $U\leq4.0$, 15 iterations for $4.0<U\leq5.0$, 25 iterations for $5.0<U<6.0$, and 35 iterations for $U\geq6.0$, producing a variety of intermediate hybridization functions along the convergence trajectory. 
To construct the dataset, we retain iterations for which the KL divergence between the spectra associated with $G^R$ and $\Delta^R$ is below $10^{-3}$; from each run, we take the final iteration and then every third preceding iteration. 
To further emphasize the strong-coupling regime, we include the first 12 DMFT iterations for $U\geq5.0$, retain all data points with $U\geq6.0$, and randomly sample from the remaining candidates.
The resulting training dataset contains 550 data points, of which 185 have $U\leq5.0$. We use only the real parts of the moments in this benchmark as the moments are purely real for particle-hole symmetric spectra.

For the doped benchmark, the dataset is built from converged DMFT loop histories on a $0.1$-spaced interaction grid, $U=1.0,1.1,\ldots,4.0$, including both doped single-orbital data and the half-filled branch at $\mu=0$. 
Here, $\mu$ denotes the shift relative to the half-filled chemical potential $\mu_{\mathrm{half}}=U/2$. 
We run calculations for $\mu/U=0,~1/16,~3/32,~5/32,~1/8,~3/16,$ and $1/4$, with the $\mu/U=3/32$ and $5/32$ branches included only for $U>3.0$.
All doped simulations are run for 10 iterations, and the density difference between successive iterations is at the level of $10^{-3}$.
The $\mu/U=1/8$ branch is held out as the test set, while all other available branches are used for training.
This gives 1124 training samples.

\section{Antiferromagnetic DMFT data details}
\label{app:afm_dmft}

For the antiferromagnetic benchmark, the local Green's function, hybridization function, and self-energy are resolved in both spin and sublattice indices. 
Collinear N\'eel order is imposed through the two-sublattice symmetry
\begin{align}
\mathbf{X}_{A\uparrow}(\omega) = \mathbf{X}_{B\downarrow}(\omega),
\qquad
\mathbf{X}_{A\downarrow}(\omega) = \mathbf{X}_{B\uparrow}(\omega),
\end{align}
for $\mathbf{X}=\mathbf{G},\boldsymbol{\Delta},\boldsymbol{\Sigma}$. 
At half filling, particle-hole symmetry further relates the two spin components on a given sublattice. 
For the retarded Green's function and hybridization function, this relation is
\begin{align}
\mathbf{X}^R_{A\uparrow}(\omega)
=
-\left[\mathbf{X}^R_{A\downarrow}(-\omega)\right]^*,
\qquad 
\mathbf{X}=\mathbf{G},\boldsymbol{\Delta}.
\end{align}
For the unshifted self-energy in the convention $\mu=U/2$, the corresponding relation is
\begin{align}
\boldsymbol{\Sigma}^R_{A\uparrow}(\omega)
=
U-\left[\boldsymbol{\Sigma}^R_{A\downarrow}(-\omega)\right]^* .
\end{align}
These symmetry relations are enforced throughout the DMFT loop, so only one sublattice impurity problem needs to be solved explicitly; the other is generated by symmetry.
To seed the antiferromagnetic solution, we initialize the two independent spin-sublattice components of the self-energy with a small imbalance
\begin{align}
\boldsymbol{\Sigma}^R_{A\uparrow}(\omega)
&=
0.51U+\frac{1}{\omega+0.5i},
\\
\boldsymbol{\Sigma}^R_{A\downarrow}(\omega)
&=
0.49U+\frac{1}{\omega+0.5i},
\end{align}
while the remaining components are generated from the N\'eel symmetry relations above.

At each DMFT iteration, the local lattice Green's function is obtained from the current self-energy by integrating over the cubic-lattice Brillouin zone. 
For the nearest-neighbor cubic lattice, the two-sublattice Green's function gives
\begin{align}
\mathbf{G}^R_{A\sigma}(\omega)
=
\frac{1}{N_k}
\sum_{\mathbf{k}}
\frac{\zeta_{B\sigma}(\omega)}
{
\zeta_{A\sigma}(\omega)\zeta_{B\sigma}(\omega)
-
B_{\mathbf{k}}^2
},
\label{eq:afm_dmft_lattice_green}
\end{align}
with an analogous expression for $\mathbf{G}^R_{B\sigma}(\omega)$ obtained by exchanging $A$ and $B$. 
Here,
\begin{align}
\zeta_{s\sigma}(\omega)
=
\omega+i\eta+\mu-\boldsymbol{\Sigma}^R_{s\sigma}(\omega),
\qquad
s=A,B,
\end{align}
and the inter-sublattice hopping is
\begin{align}
B_{\mathbf{k}}
=
-2t\left(\cos k_x+\cos k_y+\cos k_z\right).
\end{align}
The hybridization function is then updated from the impurity Dyson equation,
\begin{align}
\boldsymbol{\Delta}^R_{s\sigma}(\omega)
=
\omega+i\eta+\mu
-
\boldsymbol{\Sigma}^R_{s\sigma}(\omega)
-
\left[\mathbf{G}^R_{s\sigma}(\omega)\right]^{-1}.
\label{eq:afm_dmft_delta_update}
\end{align}
After the first iteration, the updated hybridization function is mixed with that from the previous iteration using the chosen DMFT mixing factor.

The resulting hybridization function defines the impurity problem for the next solver step. 
The impurity solver produces the real-frequency impurity Green's function, from which the impurity self-energy is reconstructed as
\begin{align}
\boldsymbol{\Sigma}^R_{s\sigma}(\omega)
=
\omega+i\eta+\mu
-
\boldsymbol{\Delta}^R_{s\sigma}(\omega)
-
\left[
\mathbf{G}^R_{s\sigma,\mathrm{imp}}(\omega)
\right]^{-1}.
\label{eq:afm_dmft_sigma_reconstruction}
\end{align}
This self-energy is then used as input for the next DMFT iteration. 
For each iteration, we store the lattice input Green's function, the input hybridization function, the impurity Green's function, the reconstructed self-energy, and the corresponding real-frequency grid.

We use $\eta=0.05$ in the DMFT self-consistency. 
Before reconstructing $\boldsymbol{\Sigma}^R(\omega)$, we apply an additional broadening of $0.05$ to $\boldsymbol{\Delta}^R(\omega)$ and $\mathbf{G}^R_{\mathrm{imp}}(\omega)$ to suppress sharp spikes and oscillations.
The dataset is built from AFM DMFT loops on a $0.05$-spaced interaction grid over $U\in[6,10]$. 
For the train--test split, the training set uses every other interaction value starting from $U=6.00$, i.e., $U=6.00,6.10,\ldots,10.00$, while the test set uses the interleaved values starting from $U=6.05$, i.e., $U=6.05,6.15,\ldots,9.95$. 
The self-consistency loop is run until the change in staggered magnetization $m=\frac{1}{2}
\left(m_A-m_B\right),
\,
m_s=
\langle  n_{s\uparrow}\rangle
-
\langle  n_{s\downarrow}\rangle$ is below $10^{-3}$.
After collecting the selected DMFT loop samples, the training set contains 484 samples.

\section{Two-orbital NCA calculations} \label{app:two_orb_method}

For the two-orbital matrix-valued benchmark, we use a steady-state
multi-orbital noncrossing approximation (NCA) impurity solver following
Ref.~\cite{Zemach_NCA_2024}.
The NCA is the lowest-order self-consistent skeleton approximation in
the hybridization expansion.
From a diagrammatic perspective, it includes infinitely many hybridization events through the noncrossing resummation, while neglecting diagrams with crossing hybridization lines.
The solver is formulated in real time on the Keldysh contour, and in the local many-body eigenbasis.
It is parametrized by the impurity Hamiltonian and the impurity--bath hybridization.

In the hybridization expansion, the bath degrees of freedom are traced
out, leaving the lesser and greater hybridization functions
\(\Delta^{<,>}_{\nu_1\nu_2}(t)\), see Eq.~\ref{eq:hyb}. These functions contract bath operators and connect
pairs of impurity states transitions on the Keldysh contour. 
We first introduce the single-branch restricted propagator and the
cross-branch vertex that appear in the hybridization expansion:
\begin{equation}
\begin{aligned}\mathcal{G}_{ii_{0}}(t) & =\operatorname{Tr}_{B}\left\{ \langle i|U(t)|i_{0}\rangle\rho_{B}\right\} ,\\
\mathcal{K}_{ji,j_{0}i_{0}}(t',t) & =\operatorname{Tr}_{B}\left\{ \langle i_{0}|U^{\dagger}(t')|i\rangle\langle j|U(t)|j_{0}\rangle\rho_{B}\right\} .
\end{aligned}
\end{equation}

Here \(U(t)=e^{-iHt}\), with
\(H=H_{\mathrm{loc}}+H_{\mathrm{bath}}+H_{\mathrm{hyb}}\), is the full
real-time evolution operator. In deriving the hybridization expansion,
this evolution is expanded in \(H_{\mathrm{hyb}}\). The indices \(j_0,i_0\) label the initial local  many-body eigenstate on the two Keldysh branches, while the indices \(j,i\) label the corresponding
final local states. We denote combined spin-orbital impurity indices by \(\nu_1,\nu_2\).
Matrix elements of impurity operators in the local many-body basis are
defined as
\begin{equation}
(d_{\nu})_{ij}
=
\langle i|d_{\nu}|j\rangle ,
\qquad
(d^{\dagger}_{\nu})_{ij}
=
\langle i|d^\dagger_{\nu}|j\rangle .
\end{equation}

In the steady-state formulation, the absolute contour times and initial state drop out, and the two-time vertex is rewritten in terms of the relative time
\(\Delta t=t-t'\),
\begin{equation}
\mathcal{K}_{ji,j_0 i_0}(t,t')
\xrightarrow[t,t'\to\infty]{}
\mathcal{K}_{ji}^{\mathrm{ss}}(\Delta t).
\end{equation}
Here $\Delta t$ is taken to be within a finite decorrelation timescale $t_\text{max}$ embodying a numerical parameter in the steady-state scheme.
The time $t_\text{max}$ is chosen to be large enough that the relevant restricted propagators and hybridization kernels have decayed within it.
As a result, the computational complexity scales linearly in $t_\text{max}$ rather than quadratically in the propagation time: generally a large advantage.
The cost is the need to introduce an additional self-consistency loop expressing the vertex in terms of itself.

The NCA resums an infinite subset of hybridization expansion contributions to
the propagators.
The propagator and vertex are obtained by solving their corresponding Dyson equations self-consistently.
These are, respectively,
\begin{equation}
\begin{aligned}\mathcal{G}_{i,i_{0}}(t)= & g_{ii_{0}}\left(t\right)-\sum_{i_{1}i_{2}}\int^{t}_{0}d\tau_{1}\int^{\tau_{1}}_{0}d\tau_{2}\\
 & \thinspace g_{i,i_{2}}\left(t-\tau_{1}\right)\Sigma^{\mathcal{G}}_{i_{2}i_{1}}\left(\tau_{1}-\tau_{2}\right)\mathcal{G}_{i_{1}i_{0}}\left(\tau_{2}\right),\\
\mathcal{K}^{\mathrm{ss}}_{ji}(t)= & \sum_{i_{1}i_{2}j_{1}j_{2}}\int^{t}_{-t_{\text{max}}/2}d\Lambda\int^{\Lambda+t_{\text{max}}/2}_{\Lambda}d\Delta\\
 & \thinspace\mathcal{G}^{\dagger}_{i,i_{1}}\left(t-\Lambda\right)\Sigma^{\mathcal{K}}_{j_{1}i_{1}}\left(\Delta\right)\mathcal{G}_{j,j_{1}}\left(\Delta-\Lambda\right).
\end{aligned}
\end{equation}

Here, the bare local propagator is
\begin{equation}
g_{ii_0}(t)
=
\delta_{ii_0}e^{-iE_i t},
\end{equation}
the single-branch and cross-branch NCA self-energies are
\begin{equation}
\begin{aligned}
\Sigma^{\mathcal{G}}_{i_{1}j_{1}}(t)
&\equiv
\sum_{i_{2}j_{2}}
\xi_{j_{1}i_{1},j_{2}i_{2}}(t)\,
\mathcal{G}_{i_{2}j_{2}}(t),
\\
\Sigma^{\mathcal{K}}_{j_{1}i_{1}}(t)
&\equiv
\sum_{j_{0}i_{0}}
\xi_{j_{0}i_{0},j_{1}i_{1}}(t)\,
\mathcal{K}^{\mathrm{ss}}_{j_{0}i_{0}}(t).
\end{aligned}
\label{eq:nca_self_energies}
\end{equation}
and are parametrized by the hybridization kernel
\begin{equation}
\begin{aligned}
\xi_{j_{1}i_{1},j_{2}i_{2}}(t)
= {}&
\sum_{\nu_1\nu_2}
\Bigl[
\Delta^{<}_{\nu_1\nu_2}(t)
(d_{\nu_1})_{i_1 i_2}
(d^\dagger_{\nu_2})_{j_2 j_1}
\\
&\qquad+
\Delta^{>}_{\nu_1\nu_2}(t)
(d^\dagger_{\nu_1})_{i_1 i_2}
(d_{\nu_2})_{j_2 j_1}
\Bigr].
\end{aligned}
\label{eq:nca_xi_kernel}
\end{equation}

The retarded impurity Green's function is reconstructed from the
converged NCA vertex and restricted propagator by inserting external
physical impurity operators.
We first write
\begin{equation}
\begin{aligned}
G^{R}_{\nu_{1}\nu_{2}}(\tau)
={}&
-i\Theta(\tau)
\sum_{i,j,n,m}
\mathcal{K}_{ji}^{\mathrm{ss}}(\tau)
\mathcal{G}_{nm}(\tau)
\\
&\times
\left[
(d_{\nu_{1}})_{jn}
(d^{\dagger}_{\nu_{2}})_{mi}
+
(d^{\dagger}_{\nu_{2}})_{jn}
(d_{\nu_{1}})_{mi}
\right].
\end{aligned}
\end{equation}
The two terms
correspond to the two operator orderings in the fermionic
anticommutator: the first term represents the
\(d_{\nu_1}(\tau)d^\dagger_{\nu_2}(0)\) contribution, while the second
term represents the \(d^\dagger_{\nu_2}(0)d_{\nu_1}(\tau)\)
contribution.
After Fourier transformation,
\begin{equation}
G_{\nu_1\nu_2}^R(\omega)
=
\int_0^\infty d\tau\,
e^{i(\omega)\tau}
G_{\nu_1\nu_2}^R(\tau).
\end{equation}
Finally, the matrix-valued spectral function used as the learning target is
defined as
\begin{equation}
\mathbf{A}(\omega)
=
-\frac{1}{\pi}
\operatorname{Im}\mathbf{G}^R(\omega).
\end{equation}

\section{Two-orbital NCA data details} \label{app:two_orb_data}

For the two-orbital benchmark, the input hybridization functions are generated from random matrix-valued pole expansions. 
For each sample, we randomly choose eight pole positions $\epsilon_\ell$ on the real axis in the interval $[-10,10]$ and use a fixed broadening $\eta=4$
\begin{align}
    \boldsymbol{\Delta}^{R}(\omega)
    =
    \sum_{\ell=1}^{8}
    \frac{\mathbf{W}_{\ell}}{\omega+i\eta-\epsilon_{\ell}} .
\end{align}
The matrix-valued pole weights are parameterized as
\begin{align}
    \mathbf{W}_{\ell} = N
    \begin{pmatrix}
    a_{\ell} & s_{\ell}r_{\ell}\sqrt{a_{\ell}b_{\ell}}\\
    s_{\ell}r_{\ell}\sqrt{a_{\ell}b_{\ell}} & b_{\ell}
    \end{pmatrix},
\end{align}
where $N=20$ is an overall prefactor, $a_{\ell},b_{\ell}\in[0.8,1.0]$, $r_{\ell}\in[0.3,0.5]$, and $s_{\ell}=\pm1$ is chosen randomly. 
This construction gives positive-definite pole weights, since $a_\ell>0$, $b_\ell>0$, and $r_\ell<1$. 
The parameter $r_\ell$ controls the relative strength of the off-diagonal component, while $s_\ell$ sets its sign. 
The overall hybridization scale is chosen sufficiently large so that the NCA propagators decay in time and the steady state can be reached reliably.

The steady-state NCA calculations are performed using the two-orbital Kanamori interaction defined in Eq.~\ref{eq:kanamori}. 
We set the Hund's coupling to $J=0.1U$, the inter-orbital hopping to $t'=0.05U$, and the crystal-field splitting to $\Delta_{\mathrm{cf}}=0.05U$. 
The onsite potentials are chosen as
\begin{align}
    \varepsilon_1=-\mu-\frac{\Delta_{\mathrm{cf}}}{2},
    \qquad
    \varepsilon_2=-\mu+\frac{\Delta_{\mathrm{cf}}}{2}.
\end{align}
The interaction strength $U$ is sampled from the interval $[2,8]$. 
The chemical potential is set to
\begin{align}
    \mu=\mu_{\mathrm{half}}+\delta\mu,
    \qquad
    \mu_{\mathrm{half}}=\frac{3U-5J}{2},
\end{align}
where $\mu_{\mathrm{half}}$ is the half-filling chemical potential. 
The doping offset $\delta\mu$ is randomly sampled from $[-0.05U,0.05U]$. 
The temperature is set to $T=0.1$.
We generated 2000 training and 400 validation parameter sets, retaining only those for which the impurity calculation converged. Convergence was defined by a final vertex residual below $5 \times 10^{-5}$. After filtering, the dataset contained 1571 training samples and 291 validation samples.

\bibliography{refs}

\begin{thebibliography}{72}%
\makeatletter
\providecommand \@ifxundefined [1]{%
 \@ifx{#1\undefined}
}%
\providecommand \@ifnum [1]{%
 \ifnum #1\expandafter \@firstoftwo
 \else \expandafter \@secondoftwo
 \fi
}%
\providecommand \@ifx [1]{%
 \ifx #1\expandafter \@firstoftwo
 \else \expandafter \@secondoftwo
 \fi
}%
\providecommand \natexlab [1]{#1}%
\providecommand \enquote  [1]{``#1''}%
\providecommand \bibnamefont  [1]{#1}%
\providecommand \bibfnamefont [1]{#1}%
\providecommand \citenamefont [1]{#1}%
\providecommand \href@noop [0]{\@secondoftwo}%
\providecommand \href [0]{\begingroup \@sanitize@url \@href}%
\providecommand \@href[1]{\@@startlink{#1}\@@href}%
\providecommand \@@href[1]{\endgroup#1\@@endlink}%
\providecommand \@sanitize@url [0]{\catcode `\\12\catcode `\$12\catcode
  `\&12\catcode `\#12\catcode `\^12\catcode `\_12\catcode `\%12\relax}%
\providecommand \@@startlink[1]{}%
\providecommand \@@endlink[0]{}%
\providecommand \url  [0]{\begingroup\@sanitize@url \@url }%
\providecommand \@url [1]{\endgroup\@href {#1}{\urlprefix }}%
\providecommand \urlprefix  [0]{URL }%
\providecommand \Eprint [0]{\href }%
\providecommand \doibase [0]{https://doi.org/}%
\providecommand \selectlanguage [0]{\@gobble}%
\providecommand \bibinfo  [0]{\@secondoftwo}%
\providecommand \bibfield  [0]{\@secondoftwo}%
\providecommand \translation [1]{[#1]}%
\providecommand \BibitemOpen [0]{}%
\providecommand \bibitemStop [0]{}%
\providecommand \bibitemNoStop [0]{.\EOS\space}%
\providecommand \EOS [0]{\spacefactor3000\relax}%
\providecommand \BibitemShut  [1]{\csname bibitem#1\endcsname}%
\let\auto@bib@innerbib\@empty
\bibitem [{\citenamefont {Georges}\ \emph {et~al.}(1996)\citenamefont
  {Georges}, \citenamefont {Kotliar}, \citenamefont {Krauth},\ and\
  \citenamefont {Rozenberg}}]{Georges_DMFT_1996}%
  \BibitemOpen
  \bibfield  {author} {\bibinfo {author} {\bibfnamefont {A.}~\bibnamefont
  {Georges}}, \bibinfo {author} {\bibfnamefont {G.}~\bibnamefont {Kotliar}},
  \bibinfo {author} {\bibfnamefont {W.}~\bibnamefont {Krauth}},\ and\ \bibinfo
  {author} {\bibfnamefont {M.~J.}\ \bibnamefont {Rozenberg}},\ }\bibfield
  {title} {\bibinfo {title} {Dynamical mean-field theory of strongly correlated
  fermion systems and the limit of infinite dimensions},\ }\href
  {https://doi.org/10.1103/RevModPhys.68.13} {\bibfield  {journal} {\bibinfo
  {journal} {Rev. Mod. Phys.}\ }\textbf {\bibinfo {volume} {68}},\ \bibinfo
  {pages} {13} (\bibinfo {year} {1996})}\BibitemShut {NoStop}%
\bibitem [{\citenamefont {Kotliar}\ \emph {et~al.}(2006)\citenamefont
  {Kotliar}, \citenamefont {Savrasov}, \citenamefont {Haule}, \citenamefont
  {Oudovenko}, \citenamefont {Parcollet},\ and\ \citenamefont
  {Marianetti}}]{Kotliar_DFT_DMFT_2006}%
  \BibitemOpen
  \bibfield  {author} {\bibinfo {author} {\bibfnamefont {G.}~\bibnamefont
  {Kotliar}}, \bibinfo {author} {\bibfnamefont {S.~Y.}\ \bibnamefont
  {Savrasov}}, \bibinfo {author} {\bibfnamefont {K.}~\bibnamefont {Haule}},
  \bibinfo {author} {\bibfnamefont {V.~S.}\ \bibnamefont {Oudovenko}}, \bibinfo
  {author} {\bibfnamefont {O.}~\bibnamefont {Parcollet}},\ and\ \bibinfo
  {author} {\bibfnamefont {C.~A.}\ \bibnamefont {Marianetti}},\ }\bibfield
  {title} {\bibinfo {title} {Electronic structure calculations with dynamical
  mean-field theory},\ }\href {https://doi.org/10.1103/RevModPhys.78.865}
  {\bibfield  {journal} {\bibinfo  {journal} {Rev. Mod. Phys.}\ }\textbf
  {\bibinfo {volume} {78}},\ \bibinfo {pages} {865} (\bibinfo {year}
  {2006})}\BibitemShut {NoStop}%
\bibitem [{\citenamefont {Onida}\ \emph {et~al.}(2002)\citenamefont {Onida},
  \citenamefont {Reining},\ and\ \citenamefont {Rubio}}]{Onida02}%
  \BibitemOpen
  \bibfield  {author} {\bibinfo {author} {\bibfnamefont {G.}~\bibnamefont
  {Onida}}, \bibinfo {author} {\bibfnamefont {L.}~\bibnamefont {Reining}},\
  and\ \bibinfo {author} {\bibfnamefont {A.}~\bibnamefont {Rubio}},\ }\bibfield
   {title} {\bibinfo {title} {Electronic excitations: density-functional versus
  many-body green's-function approaches},\ }\href
  {https://doi.org/10.1103/RevModPhys.74.601} {\bibfield  {journal} {\bibinfo
  {journal} {Rev. Mod. Phys.}\ }\textbf {\bibinfo {volume} {74}},\ \bibinfo
  {pages} {601} (\bibinfo {year} {2002})}\BibitemShut {NoStop}%
\bibitem [{\citenamefont {Arsenault}\ \emph {et~al.}(2014)\citenamefont
  {Arsenault}, \citenamefont {Lopez-Bezanilla}, \citenamefont {von
  Lilienfeld},\ and\ \citenamefont {Millis}}]{Arsenault_MLAIM_2014}%
  \BibitemOpen
  \bibfield  {author} {\bibinfo {author} {\bibfnamefont {L.-F.}\ \bibnamefont
  {Arsenault}}, \bibinfo {author} {\bibfnamefont {A.}~\bibnamefont
  {Lopez-Bezanilla}}, \bibinfo {author} {\bibfnamefont {O.~A.}\ \bibnamefont
  {von Lilienfeld}},\ and\ \bibinfo {author} {\bibfnamefont {A.~J.}\
  \bibnamefont {Millis}},\ }\bibfield  {title} {\bibinfo {title} {Machine
  learning for many-body physics: The case of the {Anderson} impurity model},\
  }\href {https://doi.org/10.1103/PhysRevB.90.155136} {\bibfield  {journal}
  {\bibinfo  {journal} {Phys. Rev. B}\ }\textbf {\bibinfo {volume} {90}},\
  \bibinfo {pages} {155136} (\bibinfo {year} {2014})}\BibitemShut {NoStop}%
\bibitem [{\citenamefont {Sheridan}\ \emph {et~al.}(2021)\citenamefont
  {Sheridan}, \citenamefont {Rhodes}, \citenamefont {Jamet}, \citenamefont
  {Rungger},\ and\ \citenamefont {Weber}}]{Sheridan_MLDMFT_2021}%
  \BibitemOpen
  \bibfield  {author} {\bibinfo {author} {\bibfnamefont {E.}~\bibnamefont
  {Sheridan}}, \bibinfo {author} {\bibfnamefont {C.}~\bibnamefont {Rhodes}},
  \bibinfo {author} {\bibfnamefont {F.}~\bibnamefont {Jamet}}, \bibinfo
  {author} {\bibfnamefont {I.}~\bibnamefont {Rungger}},\ and\ \bibinfo {author}
  {\bibfnamefont {C.}~\bibnamefont {Weber}},\ }\bibfield  {title} {\bibinfo
  {title} {Data-driven dynamical mean-field theory: An error-correction
  approach to solve the quantum many-body problem using machine learning},\
  }\href {https://doi.org/10.1103/PhysRevB.104.205120} {\bibfield  {journal}
  {\bibinfo  {journal} {Phys. Rev. B}\ }\textbf {\bibinfo {volume} {104}},\
  \bibinfo {pages} {205120} (\bibinfo {year} {2021})}\BibitemShut {NoStop}%
\bibitem [{\citenamefont {Lee}\ \emph {et~al.}(2025)\citenamefont {Lee},
  \citenamefont {Zhao}, \citenamefont {Booth}, \citenamefont {Ge},\ and\
  \citenamefont {Weber}}]{Lee_SCALINN_2025}%
  \BibitemOpen
  \bibfield  {author} {\bibinfo {author} {\bibfnamefont {H.}~\bibnamefont
  {Lee}}, \bibinfo {author} {\bibfnamefont {Z.}~\bibnamefont {Zhao}}, \bibinfo
  {author} {\bibfnamefont {G.~H.}\ \bibnamefont {Booth}}, \bibinfo {author}
  {\bibfnamefont {W.}~\bibnamefont {Ge}},\ and\ \bibinfo {author}
  {\bibfnamefont {C.}~\bibnamefont {Weber}},\ }\bibfield  {title} {\bibinfo
  {title} {Language-inspired machine learning approach for solving strongly
  correlated problems with dynamical mean-field theory},\ }\href
  {https://doi.org/10.1103/bk5q-pfb2} {\bibfield  {journal} {\bibinfo
  {journal} {Phys. Rev. B}\ }\textbf {\bibinfo {volume} {112}},\ \bibinfo
  {pages} {035165} (\bibinfo {year} {2025})}\BibitemShut {NoStop}%
\bibitem [{\citenamefont {Agapov}\ \emph {et~al.}(2024)\citenamefont {Agapov},
  \citenamefont {Bertomeu}, \citenamefont {Carballo}, \citenamefont {Mendl},\
  and\ \citenamefont {Sander}}]{Agapov_GreenNN_2024}%
  \BibitemOpen
  \bibfield  {author} {\bibinfo {author} {\bibfnamefont {E.}~\bibnamefont
  {Agapov}}, \bibinfo {author} {\bibfnamefont {O.}~\bibnamefont {Bertomeu}},
  \bibinfo {author} {\bibfnamefont {A.}~\bibnamefont {Carballo}}, \bibinfo
  {author} {\bibfnamefont {C.~B.}\ \bibnamefont {Mendl}},\ and\ \bibinfo
  {author} {\bibfnamefont {A.}~\bibnamefont {Sander}},\ }\href@noop {}
  {\bibinfo {title} {Predicting interacting green's functions with neural
  networks}} (\bibinfo {year} {2024}),\ \Eprint
  {https://arxiv.org/abs/2411.13644} {arXiv:2411.13644 [cond-mat.str-el]}
  \BibitemShut {NoStop}%
\bibitem [{\citenamefont {Kakizawa}\ \emph {et~al.}(2024)\citenamefont
  {Kakizawa}, \citenamefont {Terasaki},\ and\ \citenamefont
  {Shinaoka}}]{Kakizawa_PINN_AIM_2024}%
  \BibitemOpen
  \bibfield  {author} {\bibinfo {author} {\bibfnamefont {F.}~\bibnamefont
  {Kakizawa}}, \bibinfo {author} {\bibfnamefont {S.}~\bibnamefont {Terasaki}},\
  and\ \bibinfo {author} {\bibfnamefont {H.}~\bibnamefont {Shinaoka}},\
  }\href@noop {} {\bibinfo {title} {Physics-informed neural network model for
  quantum impurity problems based on lehmann representation}} (\bibinfo {year}
  {2024}),\ \Eprint {https://arxiv.org/abs/2411.18835} {arXiv:2411.18835
  [cond-mat.str-el]} \BibitemShut {NoStop}%
\bibitem [{\citenamefont {Dong}\ \emph {et~al.}(2024)\citenamefont {Dong},
  \citenamefont {Gull},\ and\ \citenamefont {Wang}}]{Dong_equivGF_2024}%
  \BibitemOpen
  \bibfield  {author} {\bibinfo {author} {\bibfnamefont {X.}~\bibnamefont
  {Dong}}, \bibinfo {author} {\bibfnamefont {E.}~\bibnamefont {Gull}},\ and\
  \bibinfo {author} {\bibfnamefont {L.}~\bibnamefont {Wang}},\ }\bibfield
  {title} {\bibinfo {title} {Equivariant neural network for {Green's} functions
  of molecules and materials},\ }\href
  {https://doi.org/10.1103/PhysRevB.109.075112} {\bibfield  {journal} {\bibinfo
   {journal} {Phys. Rev. B}\ }\textbf {\bibinfo {volume} {109}},\ \bibinfo
  {pages} {075112} (\bibinfo {year} {2024})}\BibitemShut {NoStop}%
\bibitem [{\citenamefont {Valenti}\ \emph {et~al.}(2026)\citenamefont
  {Valenti}, \citenamefont {Park}, \citenamefont {Georges}, \citenamefont
  {Millis},\ and\ \citenamefont {Parcollet}}]{Valenti_NNembedding_2026}%
  \BibitemOpen
  \bibfield  {author} {\bibinfo {author} {\bibfnamefont {A.}~\bibnamefont
  {Valenti}}, \bibinfo {author} {\bibfnamefont {I.}~\bibnamefont {Park}},
  \bibinfo {author} {\bibfnamefont {A.}~\bibnamefont {Georges}}, \bibinfo
  {author} {\bibfnamefont {A.~J.}\ \bibnamefont {Millis}},\ and\ \bibinfo
  {author} {\bibfnamefont {O.}~\bibnamefont {Parcollet}},\ }\href@noop {}
  {\bibinfo {title} {Neural-network quantum embedding solvers for correlated
  materials}} (\bibinfo {year} {2026}),\ \Eprint
  {https://arxiv.org/abs/2603.15741} {arXiv:2603.15741 [cond-mat.str-el]}
  \BibitemShut {NoStop}%
\bibitem [{\citenamefont {Zhu}\ \emph {et~al.}(2026)\citenamefont {Zhu},
  \citenamefont {Rosenberg}, \citenamefont {Huang}, \citenamefont {Bassi},
  \citenamefont {Yang},\ and\ \citenamefont {Zhang}}]{Zhu_ML_2026}%
  \BibitemOpen
  \bibfield  {author} {\bibinfo {author} {\bibfnamefont {Y.}~\bibnamefont
  {Zhu}}, \bibinfo {author} {\bibfnamefont {P.}~\bibnamefont {Rosenberg}},
  \bibinfo {author} {\bibfnamefont {Z.}~\bibnamefont {Huang}}, \bibinfo
  {author} {\bibfnamefont {H.}~\bibnamefont {Bassi}}, \bibinfo {author}
  {\bibfnamefont {C.}~\bibnamefont {Yang}},\ and\ \bibinfo {author}
  {\bibfnamefont {S.}~\bibnamefont {Zhang}},\ }\bibfield  {title} {\bibinfo
  {title} {Transformer-based operator learning framework for self-energy in
  strongly correlated systems},\ }\href {https://doi.org/10.1103/k5s2-x6hq}
  {\bibfield  {journal} {\bibinfo  {journal} {Phys. Rev. B}\ }\textbf {\bibinfo
  {volume} {113}},\ \bibinfo {pages} {245139} (\bibinfo {year}
  {2026})}\BibitemShut {NoStop}%
\bibitem [{\citenamefont {Sturm}\ \emph {et~al.}(2021)\citenamefont {Sturm},
  \citenamefont {Carbone}, \citenamefont {Lu}, \citenamefont {Weichselbaum},\
  and\ \citenamefont {Konik}}]{Sturm_MLAIM_2021}%
  \BibitemOpen
  \bibfield  {author} {\bibinfo {author} {\bibfnamefont {E.~J.}\ \bibnamefont
  {Sturm}}, \bibinfo {author} {\bibfnamefont {M.~R.}\ \bibnamefont {Carbone}},
  \bibinfo {author} {\bibfnamefont {D.}~\bibnamefont {Lu}}, \bibinfo {author}
  {\bibfnamefont {A.}~\bibnamefont {Weichselbaum}},\ and\ \bibinfo {author}
  {\bibfnamefont {R.~M.}\ \bibnamefont {Konik}},\ }\bibfield  {title} {\bibinfo
  {title} {Predicting impurity spectral functions using machine learning},\
  }\href {https://doi.org/10.1103/PhysRevB.103.245118} {\bibfield  {journal}
  {\bibinfo  {journal} {Phys. Rev. B}\ }\textbf {\bibinfo {volume} {103}},\
  \bibinfo {pages} {245118} (\bibinfo {year} {2021})}\BibitemShut {NoStop}%
\bibitem [{\citenamefont {Ren}\ \emph {et~al.}(2021)\citenamefont {Ren},
  \citenamefont {Han},\ and\ \citenamefont {Chen}}]{Ren_NRGspectralML_2021}%
  \BibitemOpen
  \bibfield  {author} {\bibinfo {author} {\bibfnamefont {X.-Y.}\ \bibnamefont
  {Ren}}, \bibinfo {author} {\bibfnamefont {R.-S.}\ \bibnamefont {Han}},\ and\
  \bibinfo {author} {\bibfnamefont {L.}~\bibnamefont {Chen}},\ }\bibfield
  {title} {\bibinfo {title} {Learning impurity spectral functions from density
  of states},\ }\href {https://doi.org/10.1088/1361-648X/ac2533} {\bibfield
  {journal} {\bibinfo  {journal} {J. Phys.: Condens. Matter}\ }\textbf
  {\bibinfo {volume} {33}},\ \bibinfo {pages} {495601} (\bibinfo {year}
  {2021})}\BibitemShut {NoStop}%
\bibitem [{\citenamefont {Liu}\ \emph {et~al.}(2024)\citenamefont {Liu},
  \citenamefont {Han},\ and\ \citenamefont
  {Chen}}]{Liu_impuritySpectrumDL_2024}%
  \BibitemOpen
  \bibfield  {author} {\bibinfo {author} {\bibfnamefont {T.}~\bibnamefont
  {Liu}}, \bibinfo {author} {\bibfnamefont {R.-S.}\ \bibnamefont {Han}},\ and\
  \bibinfo {author} {\bibfnamefont {L.}~\bibnamefont {Chen}},\ }\bibfield
  {title} {\bibinfo {title} {Prediction of impurity spectrum function by deep
  learning algorithm},\ }\href {https://doi.org/10.1088/1674-1056/ad3908}
  {\bibfield  {journal} {\bibinfo  {journal} {Chin. Phys. B}\ }\textbf
  {\bibinfo {volume} {33}},\ \bibinfo {pages} {057102} (\bibinfo {year}
  {2024})}\BibitemShut {NoStop}%
\bibitem [{\citenamefont {Miles}\ \emph {et~al.}(2021)\citenamefont {Miles},
  \citenamefont {Carbone}, \citenamefont {Sturm}, \citenamefont {Lu},
  \citenamefont {Weichselbaum}, \citenamefont {Barros},\ and\ \citenamefont
  {Konik}}]{Miles_KondoVAE_2021}%
  \BibitemOpen
  \bibfield  {author} {\bibinfo {author} {\bibfnamefont {C.}~\bibnamefont
  {Miles}}, \bibinfo {author} {\bibfnamefont {M.~R.}\ \bibnamefont {Carbone}},
  \bibinfo {author} {\bibfnamefont {E.~J.}\ \bibnamefont {Sturm}}, \bibinfo
  {author} {\bibfnamefont {D.}~\bibnamefont {Lu}}, \bibinfo {author}
  {\bibfnamefont {A.}~\bibnamefont {Weichselbaum}}, \bibinfo {author}
  {\bibfnamefont {K.}~\bibnamefont {Barros}},\ and\ \bibinfo {author}
  {\bibfnamefont {R.~M.}\ \bibnamefont {Konik}},\ }\bibfield  {title} {\bibinfo
  {title} {Machine learning of kondo physics using variational autoencoders and
  symbolic regression},\ }\href {https://doi.org/10.1103/PhysRevB.104.235111}
  {\bibfield  {journal} {\bibinfo  {journal} {Phys. Rev. B}\ }\textbf {\bibinfo
  {volume} {104}},\ \bibinfo {pages} {235111} (\bibinfo {year}
  {2021})}\BibitemShut {NoStop}%
\bibitem [{\citenamefont {Deng}\ \emph {et~al.}(2025)\citenamefont {Deng},
  \citenamefont {Lu}, \citenamefont {Cao},\ and\ \citenamefont
  {Zhong}}]{Deng_RFDMFT_2025}%
  \BibitemOpen
  \bibfield  {author} {\bibinfo {author} {\bibfnamefont {F.}~\bibnamefont
  {Deng}}, \bibinfo {author} {\bibfnamefont {Y.}~\bibnamefont {Lu}}, \bibinfo
  {author} {\bibfnamefont {X.}~\bibnamefont {Cao}},\ and\ \bibinfo {author}
  {\bibfnamefont {Z.}~\bibnamefont {Zhong}},\ }\href@noop {} {\bibinfo {title}
  {Neural network impurity solver for real-frequency dynamical mean-field
  theory}} (\bibinfo {year} {2025}),\ \Eprint
  {https://arxiv.org/abs/2511.14505} {arXiv:2511.14505 [cond-mat.str-el]}
  \BibitemShut {NoStop}%
\bibitem [{\citenamefont {Boehnke}\ \emph {et~al.}(2011)\citenamefont
  {Boehnke}, \citenamefont {Hafermann}, \citenamefont {Ferrero}, \citenamefont
  {Lechermann},\ and\ \citenamefont {Parcollet}}]{Boehnke_Legendre_2011}%
  \BibitemOpen
  \bibfield  {author} {\bibinfo {author} {\bibfnamefont {L.}~\bibnamefont
  {Boehnke}}, \bibinfo {author} {\bibfnamefont {H.}~\bibnamefont {Hafermann}},
  \bibinfo {author} {\bibfnamefont {M.}~\bibnamefont {Ferrero}}, \bibinfo
  {author} {\bibfnamefont {F.}~\bibnamefont {Lechermann}},\ and\ \bibinfo
  {author} {\bibfnamefont {O.}~\bibnamefont {Parcollet}},\ }\bibfield  {title}
  {\bibinfo {title} {Orthogonal polynomial representation of imaginary-time
  {Green's} functions},\ }\href {https://doi.org/10.1103/PhysRevB.84.075145}
  {\bibfield  {journal} {\bibinfo  {journal} {Phys. Rev. B}\ }\textbf {\bibinfo
  {volume} {84}},\ \bibinfo {pages} {075145} (\bibinfo {year}
  {2011})}\BibitemShut {NoStop}%
\bibitem [{\citenamefont {Gull}\ \emph {et~al.}(2018)\citenamefont {Gull},
  \citenamefont {Iskakov}, \citenamefont {Krivenko}, \citenamefont {Rusakov},\
  and\ \citenamefont {Zgid}}]{Gull2018}%
  \BibitemOpen
  \bibfield  {author} {\bibinfo {author} {\bibfnamefont {E.}~\bibnamefont
  {Gull}}, \bibinfo {author} {\bibfnamefont {S.}~\bibnamefont {Iskakov}},
  \bibinfo {author} {\bibfnamefont {I.}~\bibnamefont {Krivenko}}, \bibinfo
  {author} {\bibfnamefont {A.~A.}\ \bibnamefont {Rusakov}},\ and\ \bibinfo
  {author} {\bibfnamefont {D.}~\bibnamefont {Zgid}},\ }\bibfield  {title}
  {\bibinfo {title} {Chebyshev polynomial representation of imaginary-time
  response functions},\ }\href {https://doi.org/10.1103/PhysRevB.98.075127}
  {\bibfield  {journal} {\bibinfo  {journal} {Phys. Rev. B}\ }\textbf {\bibinfo
  {volume} {98}},\ \bibinfo {pages} {075127} (\bibinfo {year}
  {2018})}\BibitemShut {NoStop}%
\bibitem [{\citenamefont {Shinaoka}\ \emph {et~al.}(2017)\citenamefont
  {Shinaoka}, \citenamefont {Otsuki}, \citenamefont {Ohzeki},\ and\
  \citenamefont {Yoshimi}}]{Shinaoka_IR_2017}%
  \BibitemOpen
  \bibfield  {author} {\bibinfo {author} {\bibfnamefont {H.}~\bibnamefont
  {Shinaoka}}, \bibinfo {author} {\bibfnamefont {J.}~\bibnamefont {Otsuki}},
  \bibinfo {author} {\bibfnamefont {M.}~\bibnamefont {Ohzeki}},\ and\ \bibinfo
  {author} {\bibfnamefont {K.}~\bibnamefont {Yoshimi}},\ }\bibfield  {title}
  {\bibinfo {title} {Compressing {Green's} function using intermediate
  representation between imaginary-time and real-frequency domains},\ }\href
  {https://doi.org/10.1103/PhysRevB.96.035147} {\bibfield  {journal} {\bibinfo
  {journal} {Phys. Rev. B}\ }\textbf {\bibinfo {volume} {96}},\ \bibinfo
  {pages} {035147} (\bibinfo {year} {2017})}\BibitemShut {NoStop}%
\bibitem [{\citenamefont {Kaye}\ \emph {et~al.}(2022)\citenamefont {Kaye},
  \citenamefont {Chen},\ and\ \citenamefont {Parcollet}}]{Kaye_DLR_2022}%
  \BibitemOpen
  \bibfield  {author} {\bibinfo {author} {\bibfnamefont {J.}~\bibnamefont
  {Kaye}}, \bibinfo {author} {\bibfnamefont {K.}~\bibnamefont {Chen}},\ and\
  \bibinfo {author} {\bibfnamefont {O.}~\bibnamefont {Parcollet}},\ }\bibfield
  {title} {\bibinfo {title} {Discrete {Lehmann} representation of
  imaginary-time {Green's} functions},\ }\href
  {https://doi.org/10.1103/PhysRevB.105.235115} {\bibfield  {journal} {\bibinfo
   {journal} {Phys. Rev. B}\ }\textbf {\bibinfo {volume} {105}},\ \bibinfo
  {pages} {235115} (\bibinfo {year} {2022})}\BibitemShut {NoStop}%
\bibitem [{\citenamefont {Jarrell}\ and\ \citenamefont
  {Gubernatis}(1996)}]{Jarrell_AC_1996}%
  \BibitemOpen
  \bibfield  {author} {\bibinfo {author} {\bibfnamefont {M.}~\bibnamefont
  {Jarrell}}\ and\ \bibinfo {author} {\bibfnamefont {J.~E.}\ \bibnamefont
  {Gubernatis}},\ }\bibfield  {title} {\bibinfo {title} {Bayesian inference and
  the analytic continuation of imaginary-time quantum {Monte Carlo} data},\
  }\href {https://doi.org/10.1016/0370-1573(95)00074-7} {\bibfield  {journal}
  {\bibinfo  {journal} {Phys. Rep.}\ }\textbf {\bibinfo {volume} {269}},\
  \bibinfo {pages} {133} (\bibinfo {year} {1996})}\BibitemShut {NoStop}%
\bibitem [{\citenamefont {Fei}\ \emph {et~al.}(2021{\natexlab{a}})\citenamefont
  {Fei}, \citenamefont {Yeh},\ and\ \citenamefont
  {Gull}}]{Fei_Nevanlinna_2021}%
  \BibitemOpen
  \bibfield  {author} {\bibinfo {author} {\bibfnamefont {J.}~\bibnamefont
  {Fei}}, \bibinfo {author} {\bibfnamefont {C.-N.}\ \bibnamefont {Yeh}},\ and\
  \bibinfo {author} {\bibfnamefont {E.}~\bibnamefont {Gull}},\ }\bibfield
  {title} {\bibinfo {title} {{Nevanlinna} analytical continuation},\ }\href
  {https://doi.org/10.1103/PhysRevLett.126.056402} {\bibfield  {journal}
  {\bibinfo  {journal} {Phys. Rev. Lett.}\ }\textbf {\bibinfo {volume} {126}},\
  \bibinfo {pages} {056402} (\bibinfo {year} {2021}{\natexlab{a}})}\BibitemShut
  {NoStop}%
\bibitem [{\citenamefont {Fei}\ \emph {et~al.}(2021{\natexlab{b}})\citenamefont
  {Fei}, \citenamefont {Yeh}, \citenamefont {Zgid},\ and\ \citenamefont
  {Gull}}]{Fei_Caratheodory_2021}%
  \BibitemOpen
  \bibfield  {author} {\bibinfo {author} {\bibfnamefont {J.}~\bibnamefont
  {Fei}}, \bibinfo {author} {\bibfnamefont {C.-N.}\ \bibnamefont {Yeh}},
  \bibinfo {author} {\bibfnamefont {D.}~\bibnamefont {Zgid}},\ and\ \bibinfo
  {author} {\bibfnamefont {E.}~\bibnamefont {Gull}},\ }\bibfield  {title}
  {\bibinfo {title} {Analytical continuation of matrix-valued functions:
  {Carathéodory} formalism},\ }\href
  {https://doi.org/10.1103/PhysRevB.104.165111} {\bibfield  {journal} {\bibinfo
   {journal} {Phys. Rev. B}\ }\textbf {\bibinfo {volume} {104}},\ \bibinfo
  {pages} {165111} (\bibinfo {year} {2021}{\natexlab{b}})}\BibitemShut
  {NoStop}%
\bibitem [{\citenamefont {Shao}\ and\ \citenamefont
  {Sandvik}(2023)}]{Shao_SAC_2023}%
  \BibitemOpen
  \bibfield  {author} {\bibinfo {author} {\bibfnamefont {H.}~\bibnamefont
  {Shao}}\ and\ \bibinfo {author} {\bibfnamefont {A.~W.}\ \bibnamefont
  {Sandvik}},\ }\bibfield  {title} {\bibinfo {title} {Progress on stochastic
  analytic continuation of quantum {Monte Carlo} data},\ }\href
  {https://doi.org/10.1016/j.physrep.2022.11.002} {\bibfield  {journal}
  {\bibinfo  {journal} {Phys. Rep.}\ }\textbf {\bibinfo {volume} {1003}},\
  \bibinfo {pages} {1} (\bibinfo {year} {2023})}\BibitemShut {NoStop}%
\bibitem [{\citenamefont {Zhang}\ and\ \citenamefont
  {Gull}(2024)}]{Zhang_minipole_2024_a}%
  \BibitemOpen
  \bibfield  {author} {\bibinfo {author} {\bibfnamefont {L.}~\bibnamefont
  {Zhang}}\ and\ \bibinfo {author} {\bibfnamefont {E.}~\bibnamefont {Gull}},\
  }\bibfield  {title} {\bibinfo {title} {Minimal pole representation and
  controlled analytic continuation of matsubara response functions},\ }\href
  {https://doi.org/10.1103/PhysRevB.110.035154} {\bibfield  {journal} {\bibinfo
   {journal} {Phys. Rev. B}\ }\textbf {\bibinfo {volume} {110}},\ \bibinfo
  {pages} {035154} (\bibinfo {year} {2024})}\BibitemShut {NoStop}%
\bibitem [{\citenamefont {Kananenka}\ \emph {et~al.}(2016)\citenamefont
  {Kananenka}, \citenamefont {Welden}, \citenamefont {Lan}, \citenamefont
  {Gull},\ and\ \citenamefont {Zgid}}]{Kananenka2016}%
  \BibitemOpen
  \bibfield  {author} {\bibinfo {author} {\bibfnamefont {A.~A.}\ \bibnamefont
  {Kananenka}}, \bibinfo {author} {\bibfnamefont {A.~R.}\ \bibnamefont
  {Welden}}, \bibinfo {author} {\bibfnamefont {T.~N.}\ \bibnamefont {Lan}},
  \bibinfo {author} {\bibfnamefont {E.}~\bibnamefont {Gull}},\ and\ \bibinfo
  {author} {\bibfnamefont {D.}~\bibnamefont {Zgid}},\ }\bibfield  {title}
  {\bibinfo {title} {Efficient temperature-dependent green's function methods
  for realistic systems: Using cubic spline interpolation to approximate
  matsubara green's functions},\ }\href
  {https://doi.org/10.1021/acs.jctc.6b00178} {\bibfield  {journal} {\bibinfo
  {journal} {Journal of Chemical Theory and Computation}\ }\textbf {\bibinfo
  {volume} {12}},\ \bibinfo {pages} {2250} (\bibinfo {year}
  {2016})}\BibitemShut {NoStop}%
\bibitem [{\citenamefont {Zhang}\ \emph {et~al.}(2024)\citenamefont {Zhang},
  \citenamefont {Yu},\ and\ \citenamefont {Gull}}]{Zhang_minipole_2024_b}%
  \BibitemOpen
  \bibfield  {author} {\bibinfo {author} {\bibfnamefont {L.}~\bibnamefont
  {Zhang}}, \bibinfo {author} {\bibfnamefont {Y.}~\bibnamefont {Yu}},\ and\
  \bibinfo {author} {\bibfnamefont {E.}~\bibnamefont {Gull}},\ }\bibfield
  {title} {\bibinfo {title} {Minimal pole representation and analytic
  continuation of matrix-valued correlation functions},\ }\href
  {https://doi.org/10.1103/PhysRevB.110.235131} {\bibfield  {journal} {\bibinfo
   {journal} {Phys. Rev. B}\ }\textbf {\bibinfo {volume} {110}},\ \bibinfo
  {pages} {235131} (\bibinfo {year} {2024})}\BibitemShut {NoStop}%
\bibitem [{\citenamefont {Zhang}\ \emph {et~al.}(2025)\citenamefont {Zhang},
  \citenamefont {Erpenbeck}, \citenamefont {Yu},\ and\ \citenamefont
  {Gull}}]{Zhang_minipole_2025}%
  \BibitemOpen
  \bibfield  {author} {\bibinfo {author} {\bibfnamefont {L.}~\bibnamefont
  {Zhang}}, \bibinfo {author} {\bibfnamefont {A.}~\bibnamefont {Erpenbeck}},
  \bibinfo {author} {\bibfnamefont {Y.}~\bibnamefont {Yu}},\ and\ \bibinfo
  {author} {\bibfnamefont {E.}~\bibnamefont {Gull}},\ }\bibfield  {title}
  {\bibinfo {title} {Minimal pole representation for spectral functions},\
  }\href {https://doi.org/10.1063/5.0273763} {\bibfield  {journal} {\bibinfo
  {journal} {The Journal of Chemical Physics}\ }\textbf {\bibinfo {volume}
  {162}},\ \bibinfo {pages} {214111} (\bibinfo {year} {2025})}\BibitemShut
  {NoStop}%
\bibitem [{\citenamefont {Erpenbeck}\ \emph {et~al.}(2026)\citenamefont
  {Erpenbeck}, \citenamefont {Zhu}, \citenamefont {Yu}, \citenamefont {Zhang},
  \citenamefont {Gerum}, \citenamefont {Goulko}, \citenamefont {Yang},
  \citenamefont {Cohen},\ and\ \citenamefont {Gull}}]{Erpenbeck26}%
  \BibitemOpen
  \bibfield  {author} {\bibinfo {author} {\bibfnamefont {A.}~\bibnamefont
  {Erpenbeck}}, \bibinfo {author} {\bibfnamefont {Y.}~\bibnamefont {Zhu}},
  \bibinfo {author} {\bibfnamefont {Y.}~\bibnamefont {Yu}}, \bibinfo {author}
  {\bibfnamefont {L.}~\bibnamefont {Zhang}}, \bibinfo {author} {\bibfnamefont
  {R.}~\bibnamefont {Gerum}}, \bibinfo {author} {\bibfnamefont
  {O.}~\bibnamefont {Goulko}}, \bibinfo {author} {\bibfnamefont
  {C.}~\bibnamefont {Yang}}, \bibinfo {author} {\bibfnamefont {G.}~\bibnamefont
  {Cohen}},\ and\ \bibinfo {author} {\bibfnamefont {E.}~\bibnamefont {Gull}},\
  }\bibfield  {title} {\bibinfo {title} {Compact representation and long-time
  extrapolation of real-time data for quantum systems using the esprit
  algorithm},\ }\href {https://doi.org/10.1103/8vzv-y74m} {\bibfield  {journal}
  {\bibinfo  {journal} {Phys. Rev. B}\ }\textbf {\bibinfo {volume} {113}},\
  \bibinfo {pages} {115129} (\bibinfo {year} {2026})}\BibitemShut {NoStop}%
\bibitem [{\citenamefont {Gazizova}\ \emph {et~al.}(2024)\citenamefont
  {Gazizova}, \citenamefont {Zhang}, \citenamefont {Gull},\ and\ \citenamefont
  {LeBlanc}}]{Gazizova24}%
  \BibitemOpen
  \bibfield  {author} {\bibinfo {author} {\bibfnamefont {D.}~\bibnamefont
  {Gazizova}}, \bibinfo {author} {\bibfnamefont {L.}~\bibnamefont {Zhang}},
  \bibinfo {author} {\bibfnamefont {E.}~\bibnamefont {Gull}},\ and\ \bibinfo
  {author} {\bibfnamefont {J.~P.~F.}\ \bibnamefont {LeBlanc}},\ }\bibfield
  {title} {\bibinfo {title} {Feynman diagrammatics based on discrete pole
  representations: A path to renormalized perturbation theories},\ }\href
  {https://doi.org/10.1103/PhysRevB.110.075158} {\bibfield  {journal} {\bibinfo
   {journal} {Phys. Rev. B}\ }\textbf {\bibinfo {volume} {110}},\ \bibinfo
  {pages} {075158} (\bibinfo {year} {2024})}\BibitemShut {NoStop}%
\bibitem [{\citenamefont {Akhiezer}(1965)}]{Akhiezer_moment_1965}%
  \BibitemOpen
  \bibfield  {author} {\bibinfo {author} {\bibfnamefont {N.~I.}\ \bibnamefont
  {Akhiezer}},\ }\href@noop {} {\emph {\bibinfo {title} {The Classical Moment
  Problem and Some Related Questions in Analysis}}}\ (\bibinfo  {publisher}
  {Oliver and Boyd},\ \bibinfo {address} {Edinburgh},\ \bibinfo {year}
  {1965})\BibitemShut {NoStop}%
\bibitem [{\citenamefont {{Comanac}}(2007)}]{Comanac07}%
  \BibitemOpen
  \bibfield  {author} {\bibinfo {author} {\bibfnamefont {A.-B.}\ \bibnamefont
  {{Comanac}}},\ }\emph {\bibinfo {title} {{Dynamical mean field theory of
  correlated electron systems: New algorithms and applications to local
  observables}}},\ \href@noop {} {Ph.D. thesis},\ \bibinfo  {school} {Columbia
  University, New York} (\bibinfo {year} {2007})\BibitemShut {NoStop}%
\bibitem [{\citenamefont {Rusakov}\ \emph {et~al.}(2014)\citenamefont
  {Rusakov}, \citenamefont {Phillips},\ and\ \citenamefont {Zgid}}]{Rusakov14}%
  \BibitemOpen
  \bibfield  {author} {\bibinfo {author} {\bibfnamefont {A.~A.}\ \bibnamefont
  {Rusakov}}, \bibinfo {author} {\bibfnamefont {J.~J.}\ \bibnamefont
  {Phillips}},\ and\ \bibinfo {author} {\bibfnamefont {D.}~\bibnamefont
  {Zgid}},\ }\bibfield  {title} {\bibinfo {title} {Local hamiltonians for
  quantitative green's function embedding methods},\ }\href
  {https://doi.org/10.1063/1.4901432} {\bibfield  {journal} {\bibinfo
  {journal} {The Journal of Chemical Physics}\ }\textbf {\bibinfo {volume}
  {141}},\ \bibinfo {pages} {194105} (\bibinfo {year} {2014})}\BibitemShut
  {NoStop}%
\bibitem [{\citenamefont {Backhouse}\ and\ \citenamefont
  {Booth}(2022)}]{Backhouse22}%
  \BibitemOpen
  \bibfield  {author} {\bibinfo {author} {\bibfnamefont {O.~J.}\ \bibnamefont
  {Backhouse}}\ and\ \bibinfo {author} {\bibfnamefont {G.~H.}\ \bibnamefont
  {Booth}},\ }\bibfield  {title} {\bibinfo {title} {Constructing
  “full-frequency” spectra via moment constraints for coupled cluster
  green’s functions},\ }\href {https://doi.org/10.1021/acs.jctc.2c00670}
  {\bibfield  {journal} {\bibinfo  {journal} {Journal of Chemical Theory and
  Computation}\ }\textbf {\bibinfo {volume} {18}},\ \bibinfo {pages} {6622}
  (\bibinfo {year} {2022})},\ \Eprint
  {https://arxiv.org/abs/https://doi.org/10.1021/acs.jctc.2c00670}
  {https://doi.org/10.1021/acs.jctc.2c00670} \BibitemShut {NoStop}%
\bibitem [{\citenamefont {Farid}(2021)}]{Farid21}%
  \BibitemOpen
  \bibfield  {author} {\bibinfo {author} {\bibfnamefont {B.}~\bibnamefont
  {Farid}},\ }\href {https://arxiv.org/abs/1912.00474} {\bibinfo {title}
  {Many-body perturbation expansions without diagrams. i. normal states}}
  (\bibinfo {year} {2021}),\ \Eprint {https://arxiv.org/abs/1912.00474}
  {arXiv:1912.00474 [cond-mat.str-el]} \BibitemShut {NoStop}%
\bibitem [{\citenamefont {Abbott}\ \emph {et~al.}(2026)\citenamefont {Abbott},
  \citenamefont {Jay},\ and\ \citenamefont {Oare}}]{Abbott26}%
  \BibitemOpen
  \bibfield  {author} {\bibinfo {author} {\bibfnamefont {R.}~\bibnamefont
  {Abbott}}, \bibinfo {author} {\bibfnamefont {W.}~\bibnamefont {Jay}},\ and\
  \bibinfo {author} {\bibfnamefont {P.}~\bibnamefont {Oare}},\ }\href
  {https://arxiv.org/abs/2602.11260} {\bibinfo {title} {Moment problems and
  spectral functions}} (\bibinfo {year} {2026}),\ \Eprint
  {https://arxiv.org/abs/2602.11260} {arXiv:2602.11260 [hep-lat]} \BibitemShut
  {NoStop}%
\bibitem [{\citenamefont {Geronimus}(1946)}]{Geronimus_trigono_1946}%
  \BibitemOpen
  \bibfield  {author} {\bibinfo {author} {\bibfnamefont {J.}~\bibnamefont
  {Geronimus}},\ }\bibfield  {title} {\bibinfo {title} {On the trigonometric
  moment problem},\ }\href {https://doi.org/10.2307/1969232} {\bibfield
  {journal} {\bibinfo  {journal} {Annals of Mathematics}\ }\textbf {\bibinfo
  {volume} {47}},\ \bibinfo {pages} {742} (\bibinfo {year} {1946})}\BibitemShut
  {NoStop}%
\bibitem [{\citenamefont {Grenander}\ and\ \citenamefont
  {Szeg{\"o}}(1958)}]{Grenander_Toeplitz_1958}%
  \BibitemOpen
  \bibfield  {author} {\bibinfo {author} {\bibfnamefont {U.}~\bibnamefont
  {Grenander}}\ and\ \bibinfo {author} {\bibfnamefont {G.}~\bibnamefont
  {Szeg{\"o}}},\ }\href@noop {} {\emph {\bibinfo {title} {Toeplitz Forms and
  Their Applications}}}\ (\bibinfo  {publisher} {University of California
  Press},\ \bibinfo {address} {Berkeley},\ \bibinfo {year} {1958})\BibitemShut
  {NoStop}%
\bibitem [{\citenamefont {de~Prony}(1795)}]{Prony_Prony_1795}%
  \BibitemOpen
  \bibfield  {author} {\bibinfo {author} {\bibfnamefont {G.~R.}\ \bibnamefont
  {de~Prony}},\ }\bibfield  {title} {\bibinfo {title} {Essai experimental et
  analytique: sur les lois de la dilatabilite des fluides elastique et sur
  celles de la force expansive de la vapeur de l'eau et de la vapeur de
  l'alkool, a differentes temperatures},\ }\href@noop {} {\bibfield  {journal}
  {\bibinfo  {journal} {Journal Polytechnique ou Bulletin du Travail fait a
  l'Ecole Centrale des Travaux Publics}\ } (\bibinfo {year}
  {1795})}\BibitemShut {NoStop}%
\bibitem [{\citenamefont {Roy}\ and\ \citenamefont
  {Kailath}(1989)}]{Roy_ESPRIT_1989}%
  \BibitemOpen
  \bibfield  {author} {\bibinfo {author} {\bibfnamefont {R.}~\bibnamefont
  {Roy}}\ and\ \bibinfo {author} {\bibfnamefont {T.}~\bibnamefont {Kailath}},\
  }\bibfield  {title} {\bibinfo {title} {Esprit-estimation of signal parameters
  via rotational invariance techniques},\ }\href
  {https://doi.org/10.1109/29.32276} {\bibfield  {journal} {\bibinfo  {journal}
  {IEEE Transactions on acoustics, speech, and signal processing}\ }\textbf
  {\bibinfo {volume} {37}},\ \bibinfo {pages} {984} (\bibinfo {year}
  {1989})}\BibitemShut {NoStop}%
\bibitem [{\citenamefont {Potts}\ and\ \citenamefont
  {Tasche}(2013)}]{Potts_ESPRIT_2013}%
  \BibitemOpen
  \bibfield  {author} {\bibinfo {author} {\bibfnamefont {D.}~\bibnamefont
  {Potts}}\ and\ \bibinfo {author} {\bibfnamefont {M.}~\bibnamefont {Tasche}},\
  }\bibfield  {title} {\bibinfo {title} {Parameter estimation for nonincreasing
  exponential sums by prony-like methods},\ }\href
  {https://doi.org/10.1016/j.laa.2012.10.036} {\bibfield  {journal} {\bibinfo
  {journal} {Linear Algebra and its Applications}\ }\textbf {\bibinfo {volume}
  {439}},\ \bibinfo {pages} {1024} (\bibinfo {year} {2013})},\ \bibinfo {note}
  {17th Conference of the International Linear Algebra Society, Braunschweig,
  Germany, August 2011}\BibitemShut {NoStop}%
\bibitem [{\citenamefont {Ying}(2022{\natexlab{a}})}]{Ying_pole_2022a}%
  \BibitemOpen
  \bibfield  {author} {\bibinfo {author} {\bibfnamefont {L.}~\bibnamefont
  {Ying}},\ }\bibfield  {title} {\bibinfo {title} {Pole recovery from noisy
  data on imaginary axis},\ }\href {https://doi.org/10.1007/s10915-022-01963-z}
  {\bibfield  {journal} {\bibinfo  {journal} {Journal of Scientific Computing}\
  }\textbf {\bibinfo {volume} {92}},\ \bibinfo {pages} {107} (\bibinfo {year}
  {2022}{\natexlab{a}})}\BibitemShut {NoStop}%
\bibitem [{\citenamefont {Ying}(2022{\natexlab{b}})}]{Ying_pole_2022b}%
  \BibitemOpen
  \bibfield  {author} {\bibinfo {author} {\bibfnamefont {L.}~\bibnamefont
  {Ying}},\ }\bibfield  {title} {\bibinfo {title} {Analytic continuation from
  limited noisy matsubara data},\ }\href
  {https://doi.org/10.1016/j.jcp.2022.111549} {\bibfield  {journal} {\bibinfo
  {journal} {Journal of Computational Physics}\ }\textbf {\bibinfo {volume}
  {469}},\ \bibinfo {pages} {111549} (\bibinfo {year}
  {2022}{\natexlab{b}})}\BibitemShut {NoStop}%
\bibitem [{\citenamefont {Kemper}\ \emph {et~al.}(2024)\citenamefont {Kemper},
  \citenamefont {Yang},\ and\ \citenamefont {Gull}}]{Kemper_Denoise_2024}%
  \BibitemOpen
  \bibfield  {author} {\bibinfo {author} {\bibfnamefont {A.~F.}\ \bibnamefont
  {Kemper}}, \bibinfo {author} {\bibfnamefont {C.}~\bibnamefont {Yang}},\ and\
  \bibinfo {author} {\bibfnamefont {E.}~\bibnamefont {Gull}},\ }\bibfield
  {title} {\bibinfo {title} {Denoising and extension of response functions in
  the time domain},\ }\href {https://doi.org/10.1103/PhysRevLett.132.160403}
  {\bibfield  {journal} {\bibinfo  {journal} {Phys. Rev. Lett.}\ }\textbf
  {\bibinfo {volume} {132}},\ \bibinfo {pages} {160403} (\bibinfo {year}
  {2024})}\BibitemShut {NoStop}%
\bibitem [{\citenamefont {Dette}\ and\ \citenamefont
  {Wagener}(2010)}]{DETTE_2010}%
  \BibitemOpen
  \bibfield  {author} {\bibinfo {author} {\bibfnamefont {H.}~\bibnamefont
  {Dette}}\ and\ \bibinfo {author} {\bibfnamefont {J.}~\bibnamefont
  {Wagener}},\ }\bibfield  {title} {\bibinfo {title} {Matrix measures on the
  unit circle, moment spaces, orthogonal polynomials and the geronimus
  relations},\ }\href {https://doi.org/10.1016/j.laa.2009.11.006} {\bibfield
  {journal} {\bibinfo  {journal} {Linear Algebra and its Applications}\
  }\textbf {\bibinfo {volume} {432}},\ \bibinfo {pages} {1609} (\bibinfo {year}
  {2010})}\BibitemShut {NoStop}%
\bibitem [{\citenamefont {Ephremidze}\ \emph {et~al.}(2009)\citenamefont
  {Ephremidze}, \citenamefont {Janashia},\ and\ \citenamefont
  {Lagvilava}}]{Ephremidze_2009}%
  \BibitemOpen
  \bibfield  {author} {\bibinfo {author} {\bibfnamefont {L.}~\bibnamefont
  {Ephremidze}}, \bibinfo {author} {\bibfnamefont {G.}~\bibnamefont
  {Janashia}},\ and\ \bibinfo {author} {\bibfnamefont {E.}~\bibnamefont
  {Lagvilava}},\ }\bibfield  {title} {\bibinfo {title} {A simple proof of the
  matrix-valued fejér-riesz theorem},\ }\href
  {https://doi.org/10.1007/s00041-008-9051-z} {\bibfield  {journal} {\bibinfo
  {journal} {Journal of Fourier Analysis and Applications}\ }\textbf {\bibinfo
  {volume} {15}},\ \bibinfo {pages} {124} (\bibinfo {year} {2009})}\BibitemShut
  {NoStop}%
\bibitem [{\citenamefont {Simon}(2005)}]{Simon_2005}%
  \BibitemOpen
  \bibfield  {author} {\bibinfo {author} {\bibfnamefont {B.}~\bibnamefont
  {Simon}},\ }\href@noop {} {\emph {\bibinfo {title} {Orthogonal Polynomials on
  the Unit Circle, Part 1: Classical Theory}}},\ \bibinfo {series} {American
  Mathematical Society Colloquium Publications}, Vol.~\bibinfo {volume} {54}\
  (\bibinfo  {publisher} {American Mathematical Society},\ \bibinfo {address}
  {Providence, RI},\ \bibinfo {year} {2005})\BibitemShut {NoStop}%
\bibitem [{\citenamefont {Rubtsov}\ \emph {et~al.}(2005)\citenamefont
  {Rubtsov}, \citenamefont {Savkin},\ and\ \citenamefont
  {Lichtenstein}}]{Rubtsov05}%
  \BibitemOpen
  \bibfield  {author} {\bibinfo {author} {\bibfnamefont {A.~N.}\ \bibnamefont
  {Rubtsov}}, \bibinfo {author} {\bibfnamefont {V.~V.}\ \bibnamefont
  {Savkin}},\ and\ \bibinfo {author} {\bibfnamefont {A.~I.}\ \bibnamefont
  {Lichtenstein}},\ }\bibfield  {title} {\bibinfo {title} {Continuous-time
  quantum monte carlo method for fermions},\ }\href
  {https://doi.org/10.1103/PhysRevB.72.035122} {\bibfield  {journal} {\bibinfo
  {journal} {Phys. Rev. B}\ }\textbf {\bibinfo {volume} {72}},\ \bibinfo
  {pages} {035122} (\bibinfo {year} {2005})}\BibitemShut {NoStop}%
\bibitem [{\citenamefont {Werner}\ \emph {et~al.}(2006)\citenamefont {Werner},
  \citenamefont {Comanac}, \citenamefont {de' Medici}, \citenamefont {Troyer},\
  and\ \citenamefont {Millis}}]{Werner06}%
  \BibitemOpen
  \bibfield  {author} {\bibinfo {author} {\bibfnamefont {P.}~\bibnamefont
  {Werner}}, \bibinfo {author} {\bibfnamefont {A.}~\bibnamefont {Comanac}},
  \bibinfo {author} {\bibfnamefont {L.}~\bibnamefont {de' Medici}}, \bibinfo
  {author} {\bibfnamefont {M.}~\bibnamefont {Troyer}},\ and\ \bibinfo {author}
  {\bibfnamefont {A.~J.}\ \bibnamefont {Millis}},\ }\bibfield  {title}
  {\bibinfo {title} {Continuous-time solver for quantum impurity models},\
  }\href {https://doi.org/10.1103/PhysRevLett.97.076405} {\bibfield  {journal}
  {\bibinfo  {journal} {Phys. Rev. Lett.}\ }\textbf {\bibinfo {volume} {97}},\
  \bibinfo {pages} {076405} (\bibinfo {year} {2006})}\BibitemShut {NoStop}%
\bibitem [{\citenamefont {Werner}\ and\ \citenamefont
  {Millis}(2006)}]{Werner06B}%
  \BibitemOpen
  \bibfield  {author} {\bibinfo {author} {\bibfnamefont {P.}~\bibnamefont
  {Werner}}\ and\ \bibinfo {author} {\bibfnamefont {A.~J.}\ \bibnamefont
  {Millis}},\ }\bibfield  {title} {\bibinfo {title} {Hybridization expansion
  impurity solver: General formulation and application to kondo lattice and
  two-orbital models},\ }\href {https://doi.org/10.1103/PhysRevB.74.155107}
  {\bibfield  {journal} {\bibinfo  {journal} {Phys. Rev. B}\ }\textbf {\bibinfo
  {volume} {74}},\ \bibinfo {pages} {155107} (\bibinfo {year}
  {2006})}\BibitemShut {NoStop}%
\bibitem [{\citenamefont {Gull}\ \emph {et~al.}(2008)\citenamefont {Gull},
  \citenamefont {Werner}, \citenamefont {Parcollet},\ and\ \citenamefont
  {Troyer}}]{Gull08}%
  \BibitemOpen
  \bibfield  {author} {\bibinfo {author} {\bibfnamefont {E.}~\bibnamefont
  {Gull}}, \bibinfo {author} {\bibfnamefont {P.}~\bibnamefont {Werner}},
  \bibinfo {author} {\bibfnamefont {O.}~\bibnamefont {Parcollet}},\ and\
  \bibinfo {author} {\bibfnamefont {M.}~\bibnamefont {Troyer}},\ }\bibfield
  {title} {\bibinfo {title} {Continuous-time auxiliary-field monte carlo for
  quantum impurity models},\ }\href
  {https://doi.org/10.1209/0295-5075/82/57003} {\bibfield  {journal} {\bibinfo
  {journal} {Europhysics Letters}\ }\textbf {\bibinfo {volume} {82}},\ \bibinfo
  {pages} {57003} (\bibinfo {year} {2008})}\BibitemShut {NoStop}%
\bibitem [{\citenamefont {Gull}\ \emph {et~al.}(2011)\citenamefont {Gull},
  \citenamefont {Millis}, \citenamefont {Lichtenstein}, \citenamefont
  {Rubtsov}, \citenamefont {Troyer},\ and\ \citenamefont
  {Werner}}]{Gull_CTQMC_2011}%
  \BibitemOpen
  \bibfield  {author} {\bibinfo {author} {\bibfnamefont {E.}~\bibnamefont
  {Gull}}, \bibinfo {author} {\bibfnamefont {A.~J.}\ \bibnamefont {Millis}},
  \bibinfo {author} {\bibfnamefont {A.~I.}\ \bibnamefont {Lichtenstein}},
  \bibinfo {author} {\bibfnamefont {A.~N.}\ \bibnamefont {Rubtsov}}, \bibinfo
  {author} {\bibfnamefont {M.}~\bibnamefont {Troyer}},\ and\ \bibinfo {author}
  {\bibfnamefont {P.}~\bibnamefont {Werner}},\ }\bibfield  {title} {\bibinfo
  {title} {Continuous-time monte carlo methods for quantum impurity models},\
  }\href {https://doi.org/10.1103/RevModPhys.83.349} {\bibfield  {journal}
  {\bibinfo  {journal} {Rev. Mod. Phys.}\ }\textbf {\bibinfo {volume} {83}},\
  \bibinfo {pages} {349} (\bibinfo {year} {2011})}\BibitemShut {NoStop}%
\bibitem [{\citenamefont {Ganahl}\ \emph {et~al.}(2015)\citenamefont {Ganahl},
  \citenamefont {Aichhorn}, \citenamefont {Evertz}, \citenamefont
  {Thunstr\"om}, \citenamefont {Held},\ and\ \citenamefont
  {Verstraete}}]{Ganahl_mps_2015}%
  \BibitemOpen
  \bibfield  {author} {\bibinfo {author} {\bibfnamefont {M.}~\bibnamefont
  {Ganahl}}, \bibinfo {author} {\bibfnamefont {M.}~\bibnamefont {Aichhorn}},
  \bibinfo {author} {\bibfnamefont {H.~G.}\ \bibnamefont {Evertz}}, \bibinfo
  {author} {\bibfnamefont {P.}~\bibnamefont {Thunstr\"om}}, \bibinfo {author}
  {\bibfnamefont {K.}~\bibnamefont {Held}},\ and\ \bibinfo {author}
  {\bibfnamefont {F.}~\bibnamefont {Verstraete}},\ }\bibfield  {title}
  {\bibinfo {title} {Efficient dmft impurity solver using real-time dynamics
  with matrix product states},\ }\href
  {https://doi.org/10.1103/PhysRevB.92.155132} {\bibfield  {journal} {\bibinfo
  {journal} {Phys. Rev. B}\ }\textbf {\bibinfo {volume} {92}},\ \bibinfo
  {pages} {155132} (\bibinfo {year} {2015})}\BibitemShut {NoStop}%
\bibitem [{\citenamefont {Bauernfeind}\ \emph {et~al.}(2017)\citenamefont
  {Bauernfeind}, \citenamefont {Zingl}, \citenamefont {Triebl}, \citenamefont
  {Aichhorn},\ and\ \citenamefont {Evertz}}]{Bauernfeind_ftps_2017}%
  \BibitemOpen
  \bibfield  {author} {\bibinfo {author} {\bibfnamefont {D.}~\bibnamefont
  {Bauernfeind}}, \bibinfo {author} {\bibfnamefont {M.}~\bibnamefont {Zingl}},
  \bibinfo {author} {\bibfnamefont {R.}~\bibnamefont {Triebl}}, \bibinfo
  {author} {\bibfnamefont {M.}~\bibnamefont {Aichhorn}},\ and\ \bibinfo
  {author} {\bibfnamefont {H.~G.}\ \bibnamefont {Evertz}},\ }\bibfield  {title}
  {\bibinfo {title} {Fork tensor-product states: Efficient multiorbital
  real-time dmft solver},\ }\href {https://doi.org/10.1103/PhysRevX.7.031013}
  {\bibfield  {journal} {\bibinfo  {journal} {Phys. Rev. X}\ }\textbf {\bibinfo
  {volume} {7}},\ \bibinfo {pages} {031013} (\bibinfo {year}
  {2017})}\BibitemShut {NoStop}%
\bibitem [{\citenamefont {Cao}\ \emph {et~al.}(2021)\citenamefont {Cao},
  \citenamefont {Lu}, \citenamefont {Hansmann},\ and\ \citenamefont
  {Haverkort}}]{Cao_tree_2021}%
  \BibitemOpen
  \bibfield  {author} {\bibinfo {author} {\bibfnamefont {X.}~\bibnamefont
  {Cao}}, \bibinfo {author} {\bibfnamefont {Y.}~\bibnamefont {Lu}}, \bibinfo
  {author} {\bibfnamefont {P.}~\bibnamefont {Hansmann}},\ and\ \bibinfo
  {author} {\bibfnamefont {M.~W.}\ \bibnamefont {Haverkort}},\ }\bibfield
  {title} {\bibinfo {title} {Tree tensor-network real-time multiorbital
  impurity solver: Spin-orbit coupling and correlation functions in
  ${\mathrm{sr}}_{2}{\mathrm{ruo}}_{4}$},\ }\href
  {https://doi.org/10.1103/PhysRevB.104.115119} {\bibfield  {journal} {\bibinfo
   {journal} {Phys. Rev. B}\ }\textbf {\bibinfo {volume} {104}},\ \bibinfo
  {pages} {115119} (\bibinfo {year} {2021})}\BibitemShut {NoStop}%
\bibitem [{\citenamefont {Cao}\ \emph {et~al.}(2024)\citenamefont {Cao},
  \citenamefont {Lu}, \citenamefont {Stoudenmire},\ and\ \citenamefont
  {Parcollet}}]{Cao_complextime_2024}%
  \BibitemOpen
  \bibfield  {author} {\bibinfo {author} {\bibfnamefont {X.}~\bibnamefont
  {Cao}}, \bibinfo {author} {\bibfnamefont {Y.}~\bibnamefont {Lu}}, \bibinfo
  {author} {\bibfnamefont {E.~M.}\ \bibnamefont {Stoudenmire}},\ and\ \bibinfo
  {author} {\bibfnamefont {O.}~\bibnamefont {Parcollet}},\ }\bibfield  {title}
  {\bibinfo {title} {Dynamical correlation functions from complex time
  evolution},\ }\href {https://doi.org/10.1103/PhysRevB.109.235110} {\bibfield
  {journal} {\bibinfo  {journal} {Phys. Rev. B}\ }\textbf {\bibinfo {volume}
  {109}},\ \bibinfo {pages} {235110} (\bibinfo {year} {2024})}\BibitemShut
  {NoStop}%
\bibitem [{\citenamefont {Grundner}\ \emph {et~al.}(2024)\citenamefont
  {Grundner}, \citenamefont {Westhoff}, \citenamefont {Kugler}, \citenamefont
  {Parcollet},\ and\ \citenamefont {Schollw\"ock}}]{Grundner_complextime_2024}%
  \BibitemOpen
  \bibfield  {author} {\bibinfo {author} {\bibfnamefont {M.}~\bibnamefont
  {Grundner}}, \bibinfo {author} {\bibfnamefont {P.}~\bibnamefont {Westhoff}},
  \bibinfo {author} {\bibfnamefont {F.~B.}\ \bibnamefont {Kugler}}, \bibinfo
  {author} {\bibfnamefont {O.}~\bibnamefont {Parcollet}},\ and\ \bibinfo
  {author} {\bibfnamefont {U.}~\bibnamefont {Schollw\"ock}},\ }\bibfield
  {title} {\bibinfo {title} {Complex time evolution in tensor networks and
  time-dependent green's functions},\ }\href
  {https://doi.org/10.1103/PhysRevB.109.155124} {\bibfield  {journal} {\bibinfo
   {journal} {Phys. Rev. B}\ }\textbf {\bibinfo {volume} {109}},\ \bibinfo
  {pages} {155124} (\bibinfo {year} {2024})}\BibitemShut {NoStop}%
\bibitem [{\citenamefont {Yu}\ \emph {et~al.}(2026)\citenamefont {Yu},
  \citenamefont {Zhang}, \citenamefont {Gull}, \citenamefont {Cao},\ and\
  \citenamefont {Dong}}]{Yu_complex_2026}%
  \BibitemOpen
  \bibfield  {author} {\bibinfo {author} {\bibfnamefont {Y.}~\bibnamefont
  {Yu}}, \bibinfo {author} {\bibfnamefont {L.}~\bibnamefont {Zhang}}, \bibinfo
  {author} {\bibfnamefont {E.}~\bibnamefont {Gull}}, \bibinfo {author}
  {\bibfnamefont {X.}~\bibnamefont {Cao}},\ and\ \bibinfo {author}
  {\bibfnamefont {X.}~\bibnamefont {Dong}},\ }\bibfield  {title} {\bibinfo
  {title} {Multiorbital dynamical mean-field theory with a complex-time
  solver},\ }\href {https://doi.org/10.1103/yqv4-4vjx} {\bibfield  {journal}
  {\bibinfo  {journal} {Phys. Rev. Res.}\ }\textbf {\bibinfo {volume} {8}},\
  \bibinfo {pages} {023142} (\bibinfo {year} {2026})}\BibitemShut {NoStop}%
\bibitem [{\citenamefont {Eckstein}\ and\ \citenamefont
  {Werner}(2010)}]{eckstein_nonequilibrium_2010}%
  \BibitemOpen
  \bibfield  {author} {\bibinfo {author} {\bibfnamefont {M.}~\bibnamefont
  {Eckstein}}\ and\ \bibinfo {author} {\bibfnamefont {P.}~\bibnamefont
  {Werner}},\ }\bibfield  {title} {\bibinfo {title} {Nonequilibrium dynamical
  mean-field calculations based on the noncrossing approximation and its
  generalizations},\ }\href {https://doi.org/10.1103/PhysRevB.82.115115}
  {\bibfield  {journal} {\bibinfo  {journal} {Physical Review B}\ }\textbf
  {\bibinfo {volume} {82}},\ \bibinfo {pages} {115115} (\bibinfo {year}
  {2010})}\BibitemShut {NoStop}%
\bibitem [{\citenamefont {Cohen}\ \emph {et~al.}(2014)\citenamefont {Cohen},
  \citenamefont {Reichman}, \citenamefont {Millis},\ and\ \citenamefont
  {Gull}}]{cohen_greens_2014}%
  \BibitemOpen
  \bibfield  {author} {\bibinfo {author} {\bibfnamefont {G.}~\bibnamefont
  {Cohen}}, \bibinfo {author} {\bibfnamefont {D.~R.}\ \bibnamefont {Reichman}},
  \bibinfo {author} {\bibfnamefont {A.~J.}\ \bibnamefont {Millis}},\ and\
  \bibinfo {author} {\bibfnamefont {E.}~\bibnamefont {Gull}},\ }\bibfield
  {title} {\bibinfo {title} {Green's functions from real-time bold-line {Monte}
  {Carlo}},\ }\href {https://doi.org/10.1103/PhysRevB.89.115139} {\bibfield
  {journal} {\bibinfo  {journal} {Physical Review B}\ }\textbf {\bibinfo
  {volume} {89}},\ \bibinfo {pages} {115139} (\bibinfo {year}
  {2014})}\BibitemShut {NoStop}%
\bibitem [{\citenamefont {Erpenbeck}\ \emph {et~al.}(2021)\citenamefont
  {Erpenbeck}, \citenamefont {Gull},\ and\ \citenamefont
  {Cohen}}]{erpenbeck_revealing_2021}%
  \BibitemOpen
  \bibfield  {author} {\bibinfo {author} {\bibfnamefont {A.}~\bibnamefont
  {Erpenbeck}}, \bibinfo {author} {\bibfnamefont {E.}~\bibnamefont {Gull}},\
  and\ \bibinfo {author} {\bibfnamefont {G.}~\bibnamefont {Cohen}},\ }\bibfield
   {title} {\bibinfo {title} {Revealing strong correlations in higher-order
  transport statistics: {A} noncrossing approximation approach},\ }\href
  {https://doi.org/10.1103/PhysRevB.103.125431} {\bibfield  {journal} {\bibinfo
   {journal} {Physical Review B}\ }\textbf {\bibinfo {volume} {103}},\ \bibinfo
  {pages} {125431} (\bibinfo {year} {2021})}\BibitemShut {NoStop}%
\bibitem [{\citenamefont {Erpenbeck}\ and\ \citenamefont
  {Cohen}(2021)}]{erpenbeck_resolving_2021}%
  \BibitemOpen
  \bibfield  {author} {\bibinfo {author} {\bibfnamefont {A.}~\bibnamefont
  {Erpenbeck}}\ and\ \bibinfo {author} {\bibfnamefont {G.}~\bibnamefont
  {Cohen}},\ }\bibfield  {title} {\bibinfo {title} {Resolving the
  nonequilibrium kondo singlet in energy- and position-space using quantum
  measurements},\ }\href {https://doi.org/10.21468/SciPostPhys.10.6.142}
  {\bibfield  {journal} {\bibinfo  {journal} {SciPost Physics}\ }\textbf
  {\bibinfo {volume} {10}},\ \bibinfo {pages} {142} (\bibinfo {year}
  {2021})}\BibitemShut {NoStop}%
\bibitem [{\citenamefont {Zemach}\ \emph {et~al.}(2024)\citenamefont {Zemach},
  \citenamefont {Erpenbeck}, \citenamefont {Gull},\ and\ \citenamefont
  {Cohen}}]{Zemach_NCA_2024}%
  \BibitemOpen
  \bibfield  {author} {\bibinfo {author} {\bibfnamefont {I.}~\bibnamefont
  {Zemach}}, \bibinfo {author} {\bibfnamefont {A.}~\bibnamefont {Erpenbeck}},
  \bibinfo {author} {\bibfnamefont {E.}~\bibnamefont {Gull}},\ and\ \bibinfo
  {author} {\bibfnamefont {G.}~\bibnamefont {Cohen}},\ }\bibfield  {title}
  {\bibinfo {title} {Nonequilibrium steady state full counting statistics in
  the noncrossing approximation},\ }\href {https://doi.org/10.1063/5.0233876}
  {\bibfield  {journal} {\bibinfo  {journal} {The Journal of Chemical Physics}\
  }\textbf {\bibinfo {volume} {161}},\ \bibinfo {pages} {164113} (\bibinfo
  {year} {2024})}\BibitemShut {NoStop}%
\bibitem [{\citenamefont {Cohen}\ \emph {et~al.}(2015)\citenamefont {Cohen},
  \citenamefont {Gull}, \citenamefont {Reichman},\ and\ \citenamefont
  {Millis}}]{cohen_taming_2015}%
  \BibitemOpen
  \bibfield  {author} {\bibinfo {author} {\bibfnamefont {G.}~\bibnamefont
  {Cohen}}, \bibinfo {author} {\bibfnamefont {E.}~\bibnamefont {Gull}},
  \bibinfo {author} {\bibfnamefont {D.~R.}\ \bibnamefont {Reichman}},\ and\
  \bibinfo {author} {\bibfnamefont {A.~J.}\ \bibnamefont {Millis}},\ }\bibfield
   {title} {\bibinfo {title} {Taming the {Dynamical} {Sign} {Problem} in
  {Real}-{Time} {Evolution} of {Quantum} {Many}-{Body} {Problems}},\ }\href
  {https://doi.org/10.1103/PhysRevLett.115.266802} {\bibfield  {journal}
  {\bibinfo  {journal} {Physical Review Letters}\ }\textbf {\bibinfo {volume}
  {115}},\ \bibinfo {pages} {266802} (\bibinfo {year} {2015})}\BibitemShut
  {NoStop}%
\bibitem [{\citenamefont {Eidelstein}\ \emph {et~al.}(2020)\citenamefont
  {Eidelstein}, \citenamefont {Gull},\ and\ \citenamefont
  {Cohen}}]{eidelstein_multiorbital_2020}%
  \BibitemOpen
  \bibfield  {author} {\bibinfo {author} {\bibfnamefont {E.}~\bibnamefont
  {Eidelstein}}, \bibinfo {author} {\bibfnamefont {E.}~\bibnamefont {Gull}},\
  and\ \bibinfo {author} {\bibfnamefont {G.}~\bibnamefont {Cohen}},\ }\bibfield
   {title} {\bibinfo {title} {Multiorbital {Quantum} {Impurity} {Solver} for
  {General} {Interactions} and {Hybridizations}},\ }\href
  {https://doi.org/10.1103/PhysRevLett.124.206405} {\bibfield  {journal}
  {\bibinfo  {journal} {Physical Review Letters}\ }\textbf {\bibinfo {volume}
  {124}},\ \bibinfo {pages} {206405} (\bibinfo {year} {2020})}\BibitemShut
  {NoStop}%
\bibitem [{\citenamefont {Erpenbeck}\ \emph {et~al.}(2023)\citenamefont
  {Erpenbeck}, \citenamefont {Gull},\ and\ \citenamefont
  {Cohen}}]{erpenbeck_quantum_2023}%
  \BibitemOpen
  \bibfield  {author} {\bibinfo {author} {\bibfnamefont {A.}~\bibnamefont
  {Erpenbeck}}, \bibinfo {author} {\bibfnamefont {E.}~\bibnamefont {Gull}},\
  and\ \bibinfo {author} {\bibfnamefont {G.}~\bibnamefont {Cohen}},\ }\bibfield
   {title} {\bibinfo {title} {Quantum {Monte} {Carlo} {Method} in the {Steady}
  {State}},\ }\href {https://doi.org/10.1103/PhysRevLett.130.186301} {\bibfield
   {journal} {\bibinfo  {journal} {Physical Review Letters}\ }\textbf {\bibinfo
  {volume} {130}},\ \bibinfo {pages} {186301} (\bibinfo {year}
  {2023})}\BibitemShut {NoStop}%
\bibitem [{\citenamefont {Erpenbeck}\ \emph {et~al.}(2024)\citenamefont
  {Erpenbeck}, \citenamefont {Blommel}, \citenamefont {Zhang}, \citenamefont
  {Lin}, \citenamefont {Cohen},\ and\ \citenamefont
  {Gull}}]{erpenbeck_steady-state_2024}%
  \BibitemOpen
  \bibfield  {author} {\bibinfo {author} {\bibfnamefont {A.}~\bibnamefont
  {Erpenbeck}}, \bibinfo {author} {\bibfnamefont {T.}~\bibnamefont {Blommel}},
  \bibinfo {author} {\bibfnamefont {L.}~\bibnamefont {Zhang}}, \bibinfo
  {author} {\bibfnamefont {W.-T.}\ \bibnamefont {Lin}}, \bibinfo {author}
  {\bibfnamefont {G.}~\bibnamefont {Cohen}},\ and\ \bibinfo {author}
  {\bibfnamefont {E.}~\bibnamefont {Gull}},\ }\bibfield  {title} {\bibinfo
  {title} {Steady-state properties of multi-orbital systems using quantum
  {Monte} {Carlo}},\ }\href {https://doi.org/10.1063/5.0226253} {\bibfield
  {journal} {\bibinfo  {journal} {The Journal of Chemical Physics}\ }\textbf
  {\bibinfo {volume} {161}},\ \bibinfo {pages} {094104} (\bibinfo {year}
  {2024})}\BibitemShut {NoStop}%
\bibitem [{\citenamefont {Gilmer}\ \emph {et~al.}(2017)\citenamefont {Gilmer},
  \citenamefont {Schoenholz}, \citenamefont {Riley}, \citenamefont {Vinyals},\
  and\ \citenamefont {Dahl}}]{gilmer_graph_2017}%
  \BibitemOpen
  \bibfield  {author} {\bibinfo {author} {\bibfnamefont {J.}~\bibnamefont
  {Gilmer}}, \bibinfo {author} {\bibfnamefont {S.~S.}\ \bibnamefont
  {Schoenholz}}, \bibinfo {author} {\bibfnamefont {P.~F.}\ \bibnamefont
  {Riley}}, \bibinfo {author} {\bibfnamefont {O.}~\bibnamefont {Vinyals}},\
  and\ \bibinfo {author} {\bibfnamefont {G.~E.}\ \bibnamefont {Dahl}},\
  }\bibfield  {title} {\bibinfo {title} {Neural message passing for quantum
  chemistry},\ }in\ \href@noop {} {\emph {\bibinfo {booktitle} {Proceedings of
  the 34th International Conference {\allowbreak} on Machine Learning}}}\
  (\bibinfo {organization} {PMLR},\ \bibinfo {year} {2017})\ pp.\ \bibinfo
  {pages} {1263--1272}\BibitemShut {NoStop}%
\bibitem [{\citenamefont {Battaglia}\ \emph {et~al.}(2018)\citenamefont
  {Battaglia}, \citenamefont {Hamrick}, \citenamefont {Bapst}, \citenamefont
  {Sanchez-Gonzalez}, \citenamefont {Zambaldi}, \citenamefont {Malinowski},
  \citenamefont {Tacchetti}, \citenamefont {Raposo}, \citenamefont {Santoro},
  \citenamefont {Faulkner} \emph {et~al.}}]{battaglia_graph_2018}%
  \BibitemOpen
  \bibfield  {author} {\bibinfo {author} {\bibfnamefont {P.~W.}\ \bibnamefont
  {Battaglia}}, \bibinfo {author} {\bibfnamefont {J.~B.}\ \bibnamefont
  {Hamrick}}, \bibinfo {author} {\bibfnamefont {V.}~\bibnamefont {Bapst}},
  \bibinfo {author} {\bibfnamefont {A.}~\bibnamefont {Sanchez-Gonzalez}},
  \bibinfo {author} {\bibfnamefont {V.}~\bibnamefont {Zambaldi}}, \bibinfo
  {author} {\bibfnamefont {M.}~\bibnamefont {Malinowski}}, \bibinfo {author}
  {\bibfnamefont {A.}~\bibnamefont {Tacchetti}}, \bibinfo {author}
  {\bibfnamefont {D.}~\bibnamefont {Raposo}}, \bibinfo {author} {\bibfnamefont
  {A.}~\bibnamefont {Santoro}}, \bibinfo {author} {\bibfnamefont
  {R.}~\bibnamefont {Faulkner}}, \emph {et~al.},\ }\bibfield  {title} {\bibinfo
  {title} {Relational inductive biases, deep learning, and graph networks},\
  }\href@noop {} {\bibfield  {journal} {\bibinfo  {journal} {arXiv preprint
  arXiv:1806.01261}\ } (\bibinfo {year} {2018})}\BibitemShut {NoStop}%
\bibitem [{\citenamefont {Veli{\v{c}}kovi{\'c}}\ \emph
  {et~al.}(2018)\citenamefont {Veli{\v{c}}kovi{\'c}}, \citenamefont {Cucurull},
  \citenamefont {Casanova}, \citenamefont {Romero}, \citenamefont {Li{\`o}},\
  and\ \citenamefont {Bengio}}]{velickovic_graph_2018}%
  \BibitemOpen
  \bibfield  {author} {\bibinfo {author} {\bibfnamefont {P.}~\bibnamefont
  {Veli{\v{c}}kovi{\'c}}}, \bibinfo {author} {\bibfnamefont {G.}~\bibnamefont
  {Cucurull}}, \bibinfo {author} {\bibfnamefont {A.}~\bibnamefont {Casanova}},
  \bibinfo {author} {\bibfnamefont {A.}~\bibnamefont {Romero}}, \bibinfo
  {author} {\bibfnamefont {P.}~\bibnamefont {Li{\`o}}},\ and\ \bibinfo {author}
  {\bibfnamefont {Y.}~\bibnamefont {Bengio}},\ }\bibfield  {title} {\bibinfo
  {title} {Graph attention networks},\ }in\ \href
  {https://doi.org/10.48550/arXiv.1710.10903} {\emph {\bibinfo {booktitle}
  {International Conference on Learning Representations}}}\ (\bibinfo {year}
  {2018})\BibitemShut {NoStop}%
\bibitem [{\citenamefont {Perez}\ \emph {et~al.}(2018)\citenamefont {Perez},
  \citenamefont {Strub}, \citenamefont {De~Vries}, \citenamefont {Dumoulin},\
  and\ \citenamefont {Courville}}]{perez_film_2018}%
  \BibitemOpen
  \bibfield  {author} {\bibinfo {author} {\bibfnamefont {E.}~\bibnamefont
  {Perez}}, \bibinfo {author} {\bibfnamefont {F.}~\bibnamefont {Strub}},
  \bibinfo {author} {\bibfnamefont {H.}~\bibnamefont {De~Vries}}, \bibinfo
  {author} {\bibfnamefont {V.}~\bibnamefont {Dumoulin}},\ and\ \bibinfo
  {author} {\bibfnamefont {A.}~\bibnamefont {Courville}},\ }\bibfield  {title}
  {\bibinfo {title} {Film: Visual reasoning with a general conditioning
  layer},\ }in\ \href {https://doi.org/10.1609/aaai.v32i1.11671} {\emph
  {\bibinfo {booktitle} {Proceedings of the AAAI Conference on Artificial
  Intelligence}}},\ Vol.~\bibinfo {volume} {32}\ (\bibinfo {year}
  {2018})\BibitemShut {NoStop}%
\bibitem [{\citenamefont {Mead}\ and\ \citenamefont
  {Papanicolaou}(1984)}]{Mead_maxent_1984}%
  \BibitemOpen
  \bibfield  {author} {\bibinfo {author} {\bibfnamefont {L.~R.}\ \bibnamefont
  {Mead}}\ and\ \bibinfo {author} {\bibfnamefont {N.}~\bibnamefont
  {Papanicolaou}},\ }\bibfield  {title} {\bibinfo {title} {Maximum entropy in
  the problem of moments},\ }\href {https://doi.org/10.1063/1.526446}
  {\bibfield  {journal} {\bibinfo  {journal} {Journal of Mathematical Physics}\
  }\textbf {\bibinfo {volume} {25}},\ \bibinfo {pages} {2404} (\bibinfo {year}
  {1984})}\BibitemShut {NoStop}%
\end{thebibliography}%

\end{document}